\documentstyle[12pt]{article}
\newlength{\extraspace}
\newlength{\extraspaces}
\newcommand{\be}{\begin{equation}
\addtolength{\abovedisplayskip}{\extraspaces}
\addtolength{\belowdisplayskip}{\extraspaces}
\addtolength{\abovedisplayshortskip}{\extraspace}
\addtolength{\belowdisplayshortskip}{\extraspace}}
\newcommand{\ee}{\end{equation}}

\newcommand{\ba}{\begin{eqnarray}
\addtolength{\abovedisplayskip}{\extraspaces}
\addtolength{\belowdisplayskip}{\extraspaces}
\addtolength{\abovedisplayshortskip}{\extraspace}
\addtolength{\belowdisplayshortskip}{\extraspace}}
\newcommand{\ea}{\end{eqnarray}}

\newcommand{\newsection}[1]{
\vspace{15mm}
\pagebreak[3]
\addtocounter{section}{1}
\setcounter{equation}{0}
\setcounter{subsection}{0}
\setcounter{footnote}{0}
\begin{flushleft}
{\large\bf \thesection. #1}
\end{flushleft}
\nopagebreak
\medskip
\nopagebreak}

\newcommand{\Tr}{{\rm Tr}}

\begin{document}

\addtolength{\baselineskip}{.8mm}

{\thispagestyle{empty}
\noindent \hspace{1cm}  \hfill IFUP--TH/2000--27 \hspace{1cm}\\
\mbox{}                 \hfill September 2000 \hspace{1cm}\\

\begin{center}\vspace*{1.0cm}
{\large\bf Eikonal propagators and high--energy} \\
{\large\bf parton--parton scattering in gauge theories} \\
\vspace*{1.0cm}
{\large Enrico Meggiolaro}\\
\vspace*{0.5cm}{\normalsize
{Dipartimento di Fisica, \\
Universit\`a di Pisa, \\
Via Buonarroti 2, \\
I--56127 Pisa, Italy.}}\\
\vspace*{2cm}{\large \bf Abstract}
\end{center}

\noindent
In this paper we consider ``soft'' high--energy parton--parton 
scattering processes in gauge theories, i.e., elastic scattering
processes involving partons at very high squared energies $s$ in the
center of mass and small squared transferred momentum $t$ ($s \to \infty$,
$t \ll s$, typically $|t| \le 1~{\rm GeV}^2$).
By a direct resummation of perturbation theory in the limit we are
considering, we derive expressions for the truncated--connected quark
(antiquark) propagator in an external gluon field, as well as for the
residue at the pole of the full unrenormalized propagator, both for
scalar and fermion gauge theories.
These are the basic ingredients to derive high--energy parton--parton
scattering amplitudes, using the LSZ reduction formulae and a functional
integral approach. The above procedure is also extended to include
the case in which at least one of the partons is a gluon.
The meaning and the validity of the results are discussed.
}
\vfill\eject

\newsection{Introduction}

\noindent
Since the late 1950s a lot of models have been proposed to describe the
physics of hadron--hadron elastic scattering at high energies. Some of
these are ``pre--QCD'' models (see, e.g., Refs. \cite{Regge} and \cite{Yang}), 
while others are ``QCD--inspired'' models (see, e.g., Ref. \cite{Pomeron}).
With the advent of Quantum Chromo--Dynamics (QCD), which is now believed to 
be the correct theory of hadrons and their interactions, many theoretical
physicists started to study the high--energy behaviour of gauge theories
directly from their first principles.
In particular, a lot of work has been done within the framework of perturbation
theory in order to find systematic procedures for extracting the
high--energy behaviour of each amplitude and for summing these contributions
using a leading--log or eikonal approximation scheme
\cite{Cheng-Wu-book,Lipatov}.
Even if partially successful, the results obtained using these procedures
are not completely satisfactory and are not able to explain the most
relevant phenomena.

In particular, for ``soft'' high--energy scattering processes, i.e., 
elastic scattering processes at very high squared energies $s$ in the
center of mass and small squared transferred momentum $t$ ($s \to \infty$,
$t \ll s$, typically $|t| \le 1~{\rm GeV}^2$), QCD perturbation theory
cannot be safely applied, since $t$ is too small, and one has therefore
to appeal to nonperturbative QCD.
P.V. Landshoff and O. Nachtmann were maybe the first who argued
\cite{Landshoff-Nachtmann87} that the theoretical description of
measurable quantities of ``soft'' high--energy reactions (like the total
cross sections, for example) should involve in an essential way
nonperturbative QCD. Later on, Nachtmann developed a nonperturbative
analysis, based on QCD, of these ``soft'' high--energy scattering processes
\cite{Nachtmann91,Nachtmann97}: he derived formal expressions for the
quark--quark (and also quark--antiquark and antiquark--antiquark) scattering
amplitudes in the above--mentioned limit, by using a functional integral
approach and an eikonal approximation to the solution of the Dirac equation
in the presence of an external non--Abelian gauge field.

In a previous paper \cite{Meggiolaro96} we proposed an alternative
approach to high--energy quark--quark scattering based on a 
first--quantized path--integral description of quantum--field theory
developed by Fradkin in the early 1960s \cite{Fradkin}.
In this approach one obtains convenient expressions for the
truncated--connected scalar propagators in an external (gravitational,
electromagnetic, etc.) field, and the eikonal approximation can be
easily recovered in the relevant limit. Knowing the truncated--connected
propagators, one can then extract, in the manner of Lehmann, Symanzik and 
Zimmermann (LSZ) \cite{LSZ}, the scattering matrix elements in the framework
of a functional integral approach. (We remind the reader that this
method was originally adopted in Ref. \cite{Veneziano} in order to
study Planckian--energy gravitational scattering. In Ref.
\cite{Meggiolaro96} we have translated the procedure and some results
of Ref. \cite{Veneziano} to the case of Quantum Chromo--Dynamics, i.e.,
the case of quarks coupled to an external non--Abelian gauge field in
a flat space--time.)

In this paper we shall address the same problem in an even more
immediate way: by a direct resummation of perturbation theory in the 
high--energy limit that we are considering, we shall derive expressions
for the truncated--connected quark (antiquark) propagator in an
external gluon field, as well as for the residue at the pole of the
full unrenormalized propagator. (This last is also a basic ingredient
to derive parton--parton scattering amplitudes, since it appears
in the LSZ reduction formulae \cite{LSZ}.) 

The paper is organized as follows. In Sect. 2 we begin, for simplicity, 
with the case of scalar QCD, i.e., the case of a spin--0 quark 
coupled to a non--Abelian gauge field.
In Sect. 3 we extend the results to the case (more interesting from the 
physical point of view) of ``real'' fermion QCD: that is a spin--1/2 quark
coupled to a non--Abelian gauge field.
In Sect. 4 the above procedure is also extended to include
the case in which at least one of the partons is a gluon: the meaning and 
the validity of the results so obtained are discussed.
The truncated--connected propagators in an external gluon field are the basic
ingredients to derive high--energy parton--parton scattering amplitudes,
using the LSZ reduction formulae and a functional integral approach.
This was done in Ref. \cite{Meggiolaro96} and it will
be quickly reviewed in Sect. 5 for the convenience of the reader.
The results obtained in Sect. 5 are in agreement
with those of Refs. \cite{Nachtmann91,Nachtmann97,Meggiolaro96}, where
they were derived with different methods.
In Sect. 6 we give a summary of the main results, a detailed 
discussion of the approximations involved and the conclusions.

\newsection{The scattering of scalar quarks and antiquarks}

\noindent
We begin, for simplicity, with the case of scalar QCD, i.e.,
the case of a spin--0 quark (described by the scalar field $\phi$) coupled 
to a non--Abelian gauge field $A^\mu \equiv A^\mu_a T_a$, $T_a$ 
($a=1, \ldots ,N_c^2 - 1$) being the generators of the Lie algebra of the 
colour group $SU(N_c)$. We take $\phi$ to stay in the vector space of the
fundamental representation of $SU(N_c)$: i.e., $\phi$ stands for a vector
$\phi_i$ ($i = 1,\ldots, N_c$) in the colour space and $T_a$ are the
generators in the fundamental representation [$(T_a)_{ij}$, $i,j \in
\{ 1,\ldots, N_c \}$].
We limit ourselves to the case of one single flavour. The unrenormalized
Lagrangian is:
\be
L(\phi ,\phi^\dagger ,A) = 
[D^\mu \phi]^\dagger D_\mu \phi - m_0^2 \phi^\dagger \phi
- {1 \over 4} F^a_{\mu\nu} F^{a \mu\nu} ~,
\ee
where $D^\mu = \partial^\mu + ig A^\mu$ is the covariant 
derivative.

The unrenormalized scalar quark propagator will be denoted as
\be
\langle T [ \phi_i(x) \phi^\dagger_j(y) ] \rangle = S_{ij} (x-y) ~.
\label{s_prop}
\ee
Let us define the ``physical'' quark mass $m$, taken to be the pole mass,
and the residue $Z_W$ at the pole of the unrenormalized quark propagator
by the following two equations:
\be
\left[ \tilde{S} (p) \right]^{-1} \vert_{p^2 = m^2} = 0 ~~~ , ~~~
\tilde{S}_{ij} (p) \mathop\simeq_{p^2 \to m^2} 
{i Z_W ~\delta_{ij} \over p^2 - m^2 + i\varepsilon} ~,
\label{s_mass}
\ee
where $\tilde{S}_{ij} (p)$ is the unrenormalized propagator in the momentum
space:
\be
\tilde{S}_{ij} (p) \equiv \displaystyle\int d^4 z ~e^{ipz} S_{ij} (z) ~.
\label{s_tilde}
\ee
[We remind the reader that $Z_W$ is not, in general, equal to the scalar--field
renormalization constant $Z_2$ (defined as $\phi = Z^{1/2}_2 \phi_R$, where
$\phi$ is the bare field and $\phi_R$ is the renormalized field); but it is
equal to $Z_2 \tilde{z}$, where $\tilde{z}$ is the residue at the pole of
the renormalized propagator, defined as:
\be
\tilde{S}_{R ij} (p) \mathop\simeq_{p^2 \to m^2}
{i \tilde{z} ~\delta_{ij} \over p^2 - m^2 + i\varepsilon} ~.
\ee
In fact one has that:
\be
\tilde{S}_{ij} (p) = Z_2 \tilde{S}_{R ij} (p) \mathop\simeq_{p^2 \to m^2}
{i Z_2 \tilde{z} ~\delta_{ij} \over p^2 - m^2 + i\varepsilon}
\equiv {i Z_W ~\delta_{ij} \over p^2 - m^2 + i\varepsilon} ~,
\ee
that is, $Z_W = Z_2 \tilde{z}$, q.e.d. .]

In order to derive the scattering matrix 
elements following the LSZ approach \cite{LSZ}, we need to know the on--shell
{\it truncated--connected} Green functions, which are obtained from the
connected Green functions by removing the external legs calculated
on--shell. We first consider the scattering of a quark in a given
external gluon field $A^\mu$:
\be
\phi(p,j) \to \phi(p',i) ~,
\ee
where $i,~j$ are colour indices ($i,j = 1,\ldots, N_c$).
We define the truncated--connected propagator in the external 
gluon field $A^\mu$, in the momentum space as:
\be
\tilde{S}^{(tc)}_{ij} (p,p'|A) \equiv \mathop{\lim}_{p^2,p'^2 \to m^2}
{ p^2 - m^2 \over i} \tilde{S}_{ij} (p,p'|A) { p'^2 - m^2 \over i} ~,
\label{tcsp}
\ee
where $m$ is the physical mass defined above and $\tilde{S}_{ij} (p,p'|A)$ 
is the Fourier transform of $S_{ij} (x,y|A)$, the scalar propagator 
in an external gluon field, in the coordinate representation:
\be
\tilde{S}_{ij} (p,p'|A) \equiv \int d^4 x \int d^4 y \exp [i(p'x - py)]
S_{ij} (x,y|A) ~.
\label{fourier}
\ee
[Let us observe that, for convenience and simplicity reasons, we have not
included a factor $Z_W^{-2}$ in the definition (\ref{tcsp}) of the 
truncated--connected propagator. These omitted factors must be properly
included when deriving the scattering amplitudes (see Sect. 5).]
In the following we shall compute the truncated--connected propagator 
$\tilde{S}^{(tc)}_{ij} (p,p'|A)$ in the so--called {\it eikonal} approximation,
which is valid in the case of scattering particles with very high energy
($E \equiv p^0 \simeq |\vec{p}| \gg m$) and small transferred momentum
$q \equiv p' - p$ (i.e., $\sqrt{|t|} \ll E$, where $t = q^2$).
For example, if $p^\mu \simeq p'^\mu \simeq (E,E,0,0)$, one has that
$p_+ \simeq p'_+ \simeq 2E$ and $p_- \simeq p'_- \simeq 0$, where the
following general notation has been used for a given four--vector $V^\mu$:
\be
V_+ \equiv V^0 + V^1 ~~~ , ~~~ V_- \equiv V^0 - V^1 ~.
\ee
We shall call $V_+$ and $V_-$ the {\it ``longitudinal''} components of the
four--vector $V^\mu$, while $\vec{V}_\perp \equiv (V^2,V^3)$ is the component
of $V^\mu$ in the {\it ``transverse''} plane $(y,z)$.

Our strategy consists in evaluating the truncated--connected propagator
$\tilde{S}^{(tc)}_{ij} (p,p'|A)$ in each order in perturbation
theory considering $L_\phi \equiv [D^\mu \phi]^\dagger D_\mu \phi - m_0^2 
\phi^\dagger \phi = L_0 + L_{int}$, where
\be
L_0 = \partial^\mu \phi^\dagger \partial_\mu \phi - m^2 \phi^\dagger \phi
\ee
is the ``free'' (i.e., unperturbed) quark Lagrangian, which defines the
``free'' quark propagator $i/(p^2 - m^2 + i\varepsilon)$, with the
physical mass $m$, and
\be
L_{int} = -i g \left( \phi^\dagger ~\partial_\mu \phi - \partial_\mu
\phi^\dagger ~\phi \right) A^\mu
+ g^2 \phi^\dagger \phi A^\mu A_\mu
+ \delta m^2 \phi^\dagger \phi
\label{s_int}
\ee
is the ``interaction'' Lagrangian, i.e., the ``perturbation''.
The squared--mass shift $\delta m^2$ is defined as
\be
\delta m^2 \equiv m^2 - m_0^2 ~.
\ee
Let us start, therefore, by evaluating the $n$--th order term ($n \ge 1$)
in the perturbative expansion of the truncated--connected scalar propagator
in an external gluon field $A^\mu$, in the eikonal approximation.
This contribution, that we shall indicate as
$[\tilde{S}^{(tc)}_{ij} (p,p'|A)]_{(n)}$, is
schematically represented in Fig. 1a: only the quark--quark--gluon vertex
[the first term appearing in $L_{int}$ in Eq. (\ref{s_int})] contributes to
the propagator in the eikonal limit that we are considering.
Let us discuss in detail how this approximation is justified.
The key--point (see also Ref. \cite{Cheng-Wu-book} and
references therein) is that, in the high--energy limit we are considering,
$p \simeq p'$ and quarks retain their large longitudinal momenta
during their scattering process. In other words, we are assuming that
the external gluon field has a frequency distribution, i.e., Fourier
transform, $\tilde{A}_\mu (k)$ [defined by Eq. (\ref{g_tilde}) below], such 
that the relevant phase--space region in $d^4 k$ is the one where $k$ is
negligible when compared to $p$.
In Refs. \cite{Nachtmann91,Nachtmann97}, the eikonal approximation
has been done under the hypothesis that the external gluon field
contains only a limited range of frequencies: in other words,
$A_\mu (x)$ is assumed to vary slowly on the scale set
by the wavelength of the incoming waves.
We shall come back to a detailed discussion on all these
approximations and hypotheses in Sect. 5.

\begin{table}[hbt]
\centering
\small
\setlength{\tabcolsep}{1.5pc}
\caption{The relevant vertices in scalar QCD.}
\vspace{0.3cm}
\label{tab:table1}
\begin{tabular}{rr}
\hline
$Vertex:$ & $Feynman~rule:$ \\
& \\
\hline
& \\
$\phi^\dagger_i (p_2) \phi_j (p_1) A^\mu_a$ & 
$-i g (p_1 + p_2)^\mu (T_a)_{ij}$ \\
& \\
$\phi^\dagger_i \phi_j A^\mu_a A^\nu_b$ & 
$i g^2 g^{\mu\nu} \{ T_a ,T_b \}_{ij}$ \\
& \\
$\delta m^2 \phi^\dagger_i \phi_j$ & 
$i \delta m^2 \delta_{ij}$ \\
& \\
\hline
\end{tabular}
\end{table}

One must observe (see Table 1) that the scalar theory, in addition
to the quark--quark--gluon vertex, has also two other types of vertices of 
the form $g^2 \phi^\dagger \phi A^\mu A_\mu$ (quark--quark--gluon--gluon)
and $\delta m^2 \phi^\dagger \phi$ (squared--mass shift).
Yet one can easily be convinced that the contributions 
due to these additional scalar vertices to the $n$--th order 
term in the perturbative expansion of the truncated--connected scalar 
propagator $\tilde{S}^{(tc)}_{ij} (p,p'|A)$ in the high--energy limit
are suppressed with respect to the contribution coming from the 
quark--quark--gluon couplings. This is essentially due to the fact that the
quark--quark--gluon--gluon vertex  and the squared--mass shift vertex
do not carry momentum (see Table 1). 

By virtue of the above--mentioned approximations,
the expression for $[\tilde{S}^{(tc)}_{ij} (p,p'|A)]_{(n)}$
reads as follows (see Fig. 1a):
\ba
\lefteqn{
[\tilde{S}^{(tc)}_{ij} (p,p'|A)]_{(n)} \simeq } \nonumber \\
& & \simeq \displaystyle\int {d^4 q_1 \over (2\pi)^4} \ldots
\displaystyle\int {d^4 q_n \over (2\pi)^4}
~(2\pi)^4 \delta^{(4)} (q - q_1 - \ldots - q_n) \nonumber \\
& & \times \{ [ -ig 2p^{\mu_n} \tilde{A}_{\mu_n} (q_n) ]
\ldots [ -ig 2p^{\mu_1} \tilde{A}_{\mu_1} (q_1) ] \}_{ij}
\nonumber \\
& & \times {i \over (p + q_1 + \ldots + q_{n-1})^2 - m^2 + i\varepsilon}
\ldots {i \over (p + q_1)^2 - m^2 + i\varepsilon} ~,
\label{sn_1}
\ea
where $q = p' - p$ is the transferred momentum.
We have indicated with $\tilde{A}_\mu(k)$ the Fourier transform at 
four--momentum $k$ of the external gluon field $A_\mu(z)$ in the 
coordinate representation:
\be
\tilde{A}_\mu(k) \equiv \displaystyle\int d^4z ~e^{ikz} A_\mu(z) ~.
\label{g_tilde}
\ee
In the eikonal limit, one also finds that:
\be
(p + q_1 + \ldots + q_{n-1})^2 - m^2 \simeq 
2E (q_{1-} + \ldots + q_{(n-1)-}) ~,
\ee
and similarly for the other expressions. Therefore, Eq. (\ref{sn_1}) becomes,
using also Eq. (\ref{g_tilde}) to express the external gluon field in the
coordinate representation:
\ba
\lefteqn{
[\tilde{S}^{(tc)}_{ij} (p,p'|A)]_{(n)} \simeq } \nonumber \\
& & \simeq \displaystyle\int d^4 b_n \ldots \displaystyle\int d^4 b_1 
~e^{iq b_n} \{ \left[ -ig 2p^{\mu_n} A_{\mu_n} (b_n) \right] \ldots
\left[ -ig 2p^{\mu_1} A_{\mu_1} (b_1) \right] \}_{ij} \nonumber \\
& & \times \displaystyle\int {d^4 q_1 \over (2\pi)^4}
{i ~e^{iq_1(b_1 - b_n)} \over 2E q_{1-} + i\varepsilon} \ldots
\displaystyle\int {d^4 q_{n-1} \over (2\pi)^4}
{i ~e^{iq_{n-1}(b_{n-1} - b_n)} \over 2E (q_{1-} + \ldots + q_{(n-1)-})
+ i\varepsilon} ~.
\ea
The last integration can be easily performed and one obtains:
\ba
\lefteqn{
[\tilde{S}^{(tc)}_{ij} (p,p'|A)]_{(n)} \simeq } \nonumber \\
& & \simeq \displaystyle\int d^4 b_n \ldots \displaystyle\int d^4 b_1 
~e^{iq b_n} \{ \left[ -ig 2p^{\mu_n} A_{\mu_n} (b_n) \right] \ldots
\left[ -ig 2p^{\mu_1} A_{\mu_1} (b_1) \right] \}_{ij} \nonumber \\
& & \times {1 \over 2E} \delta^{(2)}(\vec{b}_{(n-1)\perp} - \vec{b}_{n\perp})
\delta (b_{(n-1)-} - b_{n-}) \theta (b_{n+} - b_{(n-1)+}) \nonumber \\
& & \times \displaystyle\int {d^4 q_1 \over (2\pi)^4}
{i ~e^{iq_1(b_1 - b_{n-1})} \over 2E q_{1-} + i\varepsilon} \ldots
\displaystyle\int {d^4 q_{n-2} \over (2\pi)^4}
{i ~e^{iq_{n-2}(b_{n-2} - b_{n-1})} \over 2E (q_{1-} + \ldots + q_{(n-2)-})
+ i\varepsilon} ~.
\ea
In the derivation of this result, we have used the following
integral expression for the {\it step} function [$\theta(\alpha)
= 1$, for $\alpha > 0$; $\theta(\alpha) = 0$, for $\alpha < 0$]:
\be
\theta (\alpha) = {1 \over 2\pi i} \displaystyle\int_{-\infty}^{+\infty}
{e^{i\alpha \omega} \over \omega - i\varepsilon} d\omega ~.
\ee
Therefore, if we proceed recursively, we obtain the following result:
\ba
\lefteqn{
[\tilde{S}^{(tc)}_{ij} (p,p'|A)]_{(n)} \simeq } \nonumber \\
& & \simeq \displaystyle\int d^2 \vec{b}_{n\perp} \displaystyle\int d b_{n-}
\displaystyle\int d b_{n+} \displaystyle\int d b_{(n-1)+} \ldots
\displaystyle\int d b_{1+} ~e^{iq b_n} \nonumber \\
& & \times {1 \over (2E)^{n-1}} \theta (b_{n+} - b_{(n-1)+}) \ldots
\theta (b_{2+} - b_{1+}) \nonumber \\
& & \times \{ \left[ -ig p^{\mu_n} A_{\mu_n} (b_n) \right] \ldots 
\left[ -ig p^{\mu_1} A_{\mu_1} (b_1) \right] \}_{ij} \vert_{ b_{i-} = b_{n-}
~; ~\vec{b}_{i\perp} = \vec{b}_{n\perp} } ~.
\label{sn_2}
\ea
Obviously, we can parametrize the coordinates $b_i$ ($i = 1, \ldots, n$)
in the integral as follows:
\be
b_i = b + p\tau_i ~~~~ (i = 1, \ldots, n) ~,
\ee
where $\tau_i$ (or, better, $\nu_i \equiv m \tau_i$) are proper--time
variables and $b$ is a four--vector with $b_- = b_{n-}$, $\vec{b}_\perp =
\vec{b}_{n\perp}$ and a fixed $b_+$. In fact, using the fact that
$p \simeq (E,E,0,0)$, i.e., $p_- \simeq 0$ and $\vec{p}_\perp \simeq 
\vec{0}_\perp$, we find that $b_{i-} = b_-$ and $\vec{b}_{i\perp} =
\vec{b}_\perp$, $\forall i = 1, \ldots, n$.
Moreover, $q = p' - p$, so that $q_+, q_- \simeq 0$ and
$q b_n \simeq -\vec{q}_\perp \cdot \vec{b}_{n\perp} =
-\vec{q}_\perp \cdot \vec{b}_\perp \simeq q b$.
Therefore, using also the fact that $d b_{i+} = p_+ d\tau_i =
2E d\tau_i$, we can re-write Eq. (\ref{sn_2}) as follows:
\ba
\lefteqn{
[\tilde{S}^{(tc)}_{ij} (p,p'|A)]_{(n)} \simeq } \nonumber \\
& & \simeq 2E \displaystyle\int [d^3 b] ~e^{iq b}
\displaystyle\int d\tau_1 \ldots \displaystyle\int d\tau_n
~\theta (\tau_n - \tau_{n-1}) \ldots \theta (\tau_2 - \tau_1)
\nonumber \\
& & \times \{ \left[ -ig p^{\mu_n} A_{\mu_n} (b + p\tau_n) \right] \ldots 
\left[ -ig p^{\mu_1} A_{\mu_1} (b + p\tau_1) \right] \}_{ij} ~,
\label{sn_3}
\ea
where we have used the notation:
\be
[d^3 b] \equiv d^2 \vec{b}_\perp d b_- ~.
\ee
As a general rule, in ``$[d^3 b]$'' one must not include the longitudinal
component of $b^\mu$ which is parallel to $p^\mu$. In other words,
if $p^\mu \simeq p'^\mu \simeq (E,E,0,0)$ (i.e., $p_+ \simeq 2E$,
$p_- \simeq 0$), one has that $[d^3 b] \equiv d^2 \vec{b}_\perp d b_-$, 
while, if $p^\mu \simeq p'^\mu \simeq (E,-E,0,0)$ (i.e., $p_+ \simeq 0$,
$p_- \simeq 2E$), then $[d^3 b] \equiv d^2 \vec{b}_\perp d b_+$.

Eq. (\ref{sn_3}) is the $n$--th order term of $\tilde{S}^{(tc)}_{ij} (p,p'|A)$.
Summing all orders ($n \ge 1$), we finally obtain (see also Ref.
\cite{Meggiolaro96}):
\ba
\lefteqn{
\tilde{S}^{(tc)}_{ij} (p,p'|A) \simeq } \nonumber \\
& & \simeq 2E \displaystyle\int [d^3 b] ~e^{iqb}
\left[ {T} \exp \left( -ig \displaystyle\int_{-\infty}^{+\infty}
A_\mu (b + p\tau) p^\mu d\tau \right) - {\bf 1} \right]_{ij} \nonumber \\
& & = 2E \displaystyle\int [d^3 b] ~e^{iqb} [W_p (b) - {\bf 1}]_{ij} ~,
\label{s_fin}
\ea
where $W_p (b) = T \exp (\ldots)$ is the {\it time}--ordered exponential,
defined as:
\ba
\lefteqn{
W_p (b) \equiv T \exp \left( -ig \displaystyle\int_{-\infty}^{+\infty}
A_\mu (b + p\tau) p^\mu d\tau \right) \equiv } \nonumber \\
& & \equiv \displaystyle\sum_{n = 0}^{\infty}
\displaystyle\int d\tau_1 \ldots \displaystyle\int d\tau_n
~\theta (\tau_n - \tau_{n-1}) \ldots \theta (\tau_2 - \tau_1)
\nonumber \\
& & \times \left[ -ig p^{\mu_n} A_{\mu_n} (b + p\tau_n) \right] \ldots
\left[ -ig p^{\mu_1} A_{\mu_1} (b + p\tau_1) \right] ~.
\label{w_def}
\ea
Eq. (\ref{s_fin}) gives the expression for the truncated--connected scalar
propagator in an external gluon field, in the eikonal approximation.

Proceeding exactly in the same way, we can also derive the following
expression for the $n$--th order term ($n \ge 1$) in the perturbative
expansion of the truncated--connected propagator of a scalar antiquark
in an external gluon field $A^\mu$,
$\phi^\dagger (p,j) \to \phi^\dagger (p',i)$, in the eikonal approximation:
\ba
\lefteqn{
[\tilde{S}^{(tc)} (\phi^\dagger_{p,j} \to \phi^\dagger_{p',i}|A)]_{(n)} 
\simeq } \nonumber \\
& & \simeq 2E \displaystyle\int [d^3 b] ~e^{iq b}
\displaystyle\int d\tau_1 \ldots \displaystyle\int d\tau_n
~\theta (\tau_n - \tau_{n-1}) \ldots \theta (\tau_2 - \tau_1)
\nonumber \\
& & \times \{ \left[ ig p^{\mu_1} A_{\mu_1} (b + p\tau_1) \right] \ldots 
\left[ ig p^{\mu_n} A_{\mu_n} (b + p\tau_n) \right] \}_{ji} ~.
\ea
And therefore, summing all orders:
\be
\tilde{S}^{(tc)} (\phi^\dagger_{p,j} \to \phi^\dagger_{p',i}|A)
\simeq 2E \displaystyle\int [d^3 b] ~e^{iqb}
\left[ \overline{T} \exp \left( ig \displaystyle\int_{-\infty}^{+\infty}
A_\mu (b + p\tau) p^\mu d\tau \right) - {\bf 1} \right]_{ji} ~.
\label{anti_s}
\ee
We have denoted with $\overline{T} \exp (\ldots)$ the {\it antitime}--ordered
exponential, defined as:
\ba
\lefteqn{
\overline{T} \exp \left( ig \displaystyle\int_{-\infty}^{+\infty}
A_\mu (b + p\tau) p^\mu d\tau \right) \equiv } \nonumber \\
& & \equiv \displaystyle\sum_{n = 0}^{\infty}
\displaystyle\int d\tau_1 \ldots \displaystyle\int d\tau_n
~\theta (\tau_n - \tau_{n-1}) \ldots \theta (\tau_2 - \tau_1)
\nonumber \\
& & \times \left[ ig p^{\mu_1} A_{\mu_1} (b + p\tau_1) \right] \ldots 
\left[ ig p^{\mu_n} A_{\mu_n} (b + p\tau_n) \right] ~.
\ea
Let us observe that in the {\it antitime} ordering the matrices $A$ are
ordered from left to right as they ``appear'' along the path going from
$\tau = -\infty$ to $\tau = +\infty$ (i.e., when increasing the proper time).
On the contrary, in the {\it time} ordering they are ordered
from left to right as they ``appear'' along the path going from $\tau =
+\infty$ to $\tau = -\infty$ (i.e., when decreasing the proper time).

Now, using the fact that $A_\mu = A^a_\mu T^a$, with $(T^a)^\dagger = T^a$,
i.e., $(T^a)_{ji} = (T^{a*})_{ij}$, we find:
\be
\left[ A_{\mu_1} (b + p\tau_1) \ldots A_{\mu_n} (b + p\tau_n) \right]_{ji} =
\left[ A^*_{\mu_n} (b + p\tau_n) \ldots A^*_{\mu_1} (b + p\tau_1) \right]_{ij}
~.
\label{cstar}
\ee
Therefore, we can write Eq. (\ref{anti_s}) as follows:
\ba
\lefteqn{
\tilde{S}^{(tc)} (\phi^\dagger_{p,j} \to \phi^\dagger_{p',i}|A)
\simeq } \nonumber \\
& & \simeq 2E \displaystyle\int [d^3 b] ~e^{iqb}
\left[ {T} \exp \left( ig \displaystyle\int_{-\infty}^{+\infty}
A^*_\mu (b + p\tau) p^\mu d\tau \right) - {\bf 1} \right]_{ij} \nonumber \\
& & \simeq 2E \displaystyle\int [d^3 b] ~e^{iqb} [W^*_p (b) - {\bf 1}]_{ij} ~,
\label{anti_s_fin}
\ea
where $W_p (b)$ has been defined in Eq. (\ref{w_def}).

When comparing with the result (\ref{s_fin}), we see that the
scattering amplitude of an antiquark in the external gluon field $A_\mu$
is equal to the scattering amplitude of a quark in the charge--conjugated
(C--transformed) gluon field $A'_\mu = -A^t_\mu = -A^*_\mu$, as expected.

\newsection{The scattering of quarks and antiquarks}

\noindent
We want now to extend the results obtained in the previous section to the 
case (more interesting from the physical point of view) of ``real'' 
fermion QCD: that is a spin--1/2 quark coupled to a non--Abelian gauge field.

As before, we limit ourselves to the case of one single flavour. 
The unrenormalized QCD Lagrangian is:
\be
L(\psi ,\psi^\dagger ,A) = 
\overline{\psi} (i\gamma^\mu D_\mu - m_0) \psi
- {1 \over 4} F^a_{\mu\nu} F^{a \mu\nu} ~,
\ee
where $D^\mu = \partial^\mu + ig A^\mu$ is the covariant 
derivative [$A^\mu \equiv A^\mu_a T_a$, $T_a$ ($a=1, \ldots ,N_c^2 - 1$) being
the generators of the Lie algebra of the colour group $SU(N_c)$]
and $\psi$ stands for a vector $\psi_i$ ($i = 1,\ldots, N_c$) in the colour
vector space of the fundamental representation.

The unrenormalized quark propagator will be denoted as
\be
\langle T [ \psi_i(x) \overline{\psi}_j(y) ] \rangle = G_{ij} (x-y) ~.
\label{f_prop}
\ee
Let us define the ``physical'' quark mass $m$, taken to be the pole mass,
and the residue $Z_W$ at the pole of the unrenormalized quark propagator
by the following two equations:
\be
\left[ \tilde{G} (p) \right]^{-1} \vert_{p^2 = m^2} = 0 ~~~ , ~~~
\tilde{G}_{ij} (p) \mathop\simeq_{p^2 \to m^2}
{i Z_W ~\delta_{ij} \over {\mathaccent 94 p} - m + i\varepsilon} ~,
\label{f_mass}
\ee
where $\tilde{G}_{ij} (p)$ is the unrenormalized propagator in the momentum
space:
\be
\tilde{G}_{ij} (p) \equiv \displaystyle\int d^4 z ~e^{ipz} G_{ij} (z) ~.
\label{f_tilde}
\ee
In Eq. (\ref{f_mass}) we have used the notation: 
${\mathaccent 94 a} \equiv \gamma^\mu a_\mu$.

As in the previous section, we consider the scattering of a quark in a
given external gluon field $A^\mu$. 
The truncated--connected fermion propagator in the momentum space
is defined as:
\be
\tilde{G}^{(tc)}_{ij} (p,p'|A) \equiv \mathop{\lim}_{p^2,p'^2 \to m^2}
{ {\mathaccent 94 p}' - m \over i} \tilde{G}_{ij} (p,p'|A)
{ {\mathaccent 94 p} - m \over i} ~,
\label{tcfp}
\ee
where $m$ is the physical quark mass defined above and
$\tilde{G}_{ij} (p,p'|A)$ is the Fourier transform [see Eq. (\ref{fourier})]
of $G_{ij} (x,y|A)$, the truncated--connected fermion propagator in
the coordinate representation.
The matrix element for the scattering of a quark in a given external
gluon field $A^\mu$,
\be
\psi(p,j,\beta) \to \psi(p',i,\alpha) ~,
\ee
where $i,~j$ are colour indices and $\alpha,\beta$ are spin indices, is given
by $\overline{u}_\alpha (p') \tilde{G}^{(tc)}_{ij} (p,p'|A) u_\beta (p)$,
where $u_\alpha (p)$ are the ``positive--energy'' spinors
[$({\mathaccent 94 p} - m) u_\alpha (p) = \overline{u}_\alpha (p)
({\mathaccent 94 p} - m) = 0$] with the usual relativistic normalization:
\be
\overline{u}_\alpha (p) \gamma^\mu u_\beta (p) = 2 p^\mu \delta_{\alpha\beta}
~~~ , ~~~ \overline{u}_\alpha (p) u_\beta (p) = 2m \delta_{\alpha\beta} ~.
\ee
In the high--energy limit we are considering, we can make the following
replacement:
\be
\overline{u}_\alpha (p') \tilde{G}^{(tc)}_{ij} (p,p'|A) u_\beta (p)
\simeq \delta_{\alpha\beta} \cdot \tilde{S}^{(tc)}_{ij} (p,p'|A) ~,
\label{f_matrix}
\ee
where $\tilde{S}^{(tc)}_{ij} (p,p'|A)$ is the truncated--connected propagator
for a scalar (i.e., spin--0) quark in the external gluon field $A^\mu$,
which was discussed in the previous section.
Let us see how this approximation is justified.
As in the scalar case, our strategy consists in evaluating the
truncated--connected propagator $\tilde{G}^{(tc)}_{ij} (p,p'|A)$, or
better the quantity (\ref{f_matrix}), in each order in perturbation theory,
considering $L_\psi \equiv \overline{\psi} (i\gamma^\mu D_\mu - m_0) \psi
= L_0 + L_{int}$, where
\be
L_0 = \overline{\psi} (i\gamma^\mu \partial_\mu - m) \psi
\ee
is the ``free'' (i.e., ``unperturbed'') quark Lagrangian, which defines
the ``free'' quark propagator $i/({\mathaccent 94 p} - m + i\varepsilon)$,
with the physical mass $m$, and
\be
L_{int} = -g \overline{\psi} \gamma^\mu A_\mu \psi
+ \delta m \overline{\psi} \psi
\ee
is the ``interaction'' Lagrangian, i.e., the ``perturbation''.
The mass shift $\delta m$ is defined as:
\be
\delta m \equiv m - m_0 ~.
\ee
Using the fact that $\overline{u} (p') ({\mathaccent 94 p}' - m) u(p) =
\overline{u} (p') ({\mathaccent 94 p} - m) u(p) = 0$ and the definition 
(\ref{tcfp}) of the truncated--connected fermion propagator, one can easily
derive the following expression for the $n$--th order term in the
perturbative expansion ($n \ge 1$):
\ba
\lefteqn{
[\overline{u}_\alpha (p') \tilde{G}^{(tc)}_{ij} (p,p'|A) u_\beta (p)]_{(n)}
= } \nonumber \\
& & = \displaystyle\int {d^4 q_1 \over (2\pi)^4} \ldots
\displaystyle\int {d^4 q_n \over (2\pi)^4}
~(2\pi)^4 \delta^{(4)} (q - q_1 - \ldots - q_n) ~N^{(ferm)}_{\alpha\beta,~ij}
(q_1, \ldots, q_n) \nonumber \\
& & \times {i \over (p + q_1 + \ldots + q_{n-1})^2 - m^2 + i\varepsilon}
\ldots {i \over (p + q_1)^2 - m^2 + i\varepsilon} ~,
\label{fn_1}
\ea
where $N^{(ferm)}_{\alpha\beta,~ij} (q_1, \ldots, q_n)$ is given by:
\ba
\lefteqn{
N^{(ferm)}_{\alpha\beta,~ij} (q_1, \ldots ,q_n) \equiv }
\nonumber \\
& & \equiv \overline{u}_\alpha (p')
\{ [ -i g \gamma^{\mu_n} \tilde{A}_{\mu_n} (q_n)
+ i \delta m ~(2\pi)^4 \delta^{(4)} (q_n) \cdot {\bf 1} ]
({\mathaccent 94 p} + {\mathaccent 94 q}_1 + \ldots +
{\mathaccent 94 q}_{n-1} + m) \nonumber \\
& & \ldots ({\mathaccent 94 p} + {\mathaccent 94 q}_1 + m) 
[ -i g \gamma^{\mu_1} \tilde{A}_{\mu_1} (q_1) + i \delta m
~(2\pi)^4 \delta^{(4)} (q_1) \cdot {\bf 1} ] \}_{ij} u_\beta (p) ~.
\label{f_num}
\ea
By virtue of the eikonal approximation, the relevant phase--space region
in the $n$--th order term of the perturbative expansion (\ref{fn_1}) has the
property that $q_i$, $m$ and $\delta m$ are negligible when compared to $p$ in
the numerator $N^{(ferm)}_{\alpha\beta,~ij} (q_1, \ldots, q_n)$.
Therefore, the numerator $N^{(ferm)}_{\alpha\beta,~ij} (q_1, \ldots, q_n)$
in Eq. (\ref{fn_1}) can be approximated as:
\be
N^{(ferm)}_{\alpha\beta,~ij} (q_1, \ldots ,q_n) \simeq
\{ -ig 2p^{\mu_n} \tilde{A}_{\mu_n} (q_n) \cdot
N^{(ferm)}_{\alpha\beta} (q_1, \ldots ,q_{n-1}) \}_{ij} ~.
\label{recursive}
\ee
Using the fact that:
\ba
N^{(ferm)}_{\alpha\beta,~ij} (q_1) & = &
\overline{u}_\alpha (p')
\{ -i g \gamma^{\mu_1} \tilde{A}_{\mu_1} (q_1)
+ i \delta m ~(2\pi)^4 \delta^{(4)} (q_1) \cdot {\bf 1} \}_{ij} u_\beta (p)
\nonumber \\
& \simeq & \delta_{\alpha\beta} \cdot
[-ig 2p^{\mu_1} \tilde{A}_{\mu_1} (q_1)]_{ij} ~,
\ea
we find, proceeding recursively from Eq. (\ref{recursive}):
\be
N^{(ferm)}_{\alpha\beta,~ij} (q_1, \ldots ,q_n)
\simeq \delta_{\alpha\beta} \cdot
\{ [-ig 2p^{\mu_n} \tilde{A}_{\mu_n} (q_n)] \ldots
[-ig 2p^{\mu_1} \tilde{A}_{\mu_1} (q_1)] \}_{ij} ~.
\label{f_num_fin}
\ee
Apart from the {\it delta} function in front, which simply reflects the
fact that fermions retain their helicities during the scattering process
in the high--energy limit, this is
exactly the term we would have expected at the numerator of the $n$--th order
term in the perturbative expansion of the truncated--connected scalar
propagator $\tilde{S}^{(tc)}_{ij} (p,p'|A)$ in the high--energy limit
[the denominators in (\ref{fn_1}) are already equal to the scalar case!].
In fact, as reported in Table 1, the factors $(-ig 2p^\mu)$
in (\ref{sn_1}) come from the quark--quark--gluon vertex
of the scalar theory in the high--energy limit (when $p \simeq p'$).
From Eqs. (\ref{fn_1}), (\ref{f_num_fin}) and (\ref{sn_1}) we derive
\be
[\overline{u}_\alpha (p') \tilde{G}^{(tc)}_{ij} (p,p'|A) u_\beta (p)]_{(n)}
\simeq \delta_{\alpha\beta} \cdot [\tilde{S}^{(tc)}_{ij} (p,p'|A)]_{(n)} ~,
\ee
for each order $n$ in the perturbative expansion, so proving Eq.
(\ref{f_matrix}). Therefore, using the result (\ref{s_fin}) derived in the
previous section, we find the following expression for the quantity
(\ref{f_matrix}) in the eikonal approximation (see also Ref.
\cite{Meggiolaro96}):
\ba
\lefteqn{
\overline{u}_\alpha (p') \tilde{G}^{(tc)}_{ij} (p,p'|A) u_\beta (p)
\simeq \delta_{\alpha\beta} \cdot \tilde{S}^{(tc)}_{ij} (p,p'|A) }
\nonumber \\
& & \simeq \delta_{\alpha\beta} \cdot 2E \displaystyle\int [d^3 b] ~e^{iqb}
\left[ {T} \exp \left( -ig \displaystyle\int_{-\infty}^{+\infty}
A_\mu (b + p\tau) p^\mu d\tau \right) - {\bf 1} \right]_{ij} \nonumber \\
& & = \delta_{\alpha\beta} \cdot 2E \displaystyle\int [d^3 b] ~e^{iqb}
[W_p (b) - {\bf 1}]_{ij} ~,
\label{f_fin}
\ea
where $W_p (b)$ is the time--ordered Wilson string along the path $x(\tau) =
b + p\tau$, $\tau \in [-\infty,+\infty]$, defined in Eq. (\ref{w_def}).

Proceeding exactly in the same way, we can also derive the following
expression for the truncated--connected propagator of an antiquark in
an external gluon field $A^\mu$,
$\overline{\psi} (p,j,\beta) \to \overline{\psi} (p',i,\alpha)$,
in the eikonal approximation:
\ba
\lefteqn{
\overline{v}_\beta (p) \tilde{G}^{(tc)} (\overline{\psi}_{p,j}
\to \overline{\psi}_{p',i}|A) v_\alpha (p') \simeq \delta_{\alpha\beta}
\cdot \tilde{S}^{(tc)} (\phi^\dagger_{p,j} \to \phi^\dagger_{p',i}|A) }
\nonumber \\
& & \simeq \delta_{\alpha\beta} \cdot 2E \displaystyle\int [d^3 b] ~e^{iqb}
\left[ {T} \exp \left( ig \displaystyle\int_{-\infty}^{+\infty}
A^*_\mu (b + p\tau) p^\mu d\tau \right) - {\bf 1} \right]_{ij} \nonumber \\
& & = \delta_{\alpha\beta} \cdot 2E \displaystyle\int [d^3 b] ~e^{iqb}
[W^*_p (b) - {\bf 1}]_{ij} ~;
\label{anti_f_fin}
\ea
$v_\alpha (p)$ are the
``negative--energy'' spinors [$({\mathaccent 94 p} + m) v_\alpha (p) =
\overline{v}_\alpha (p) ({\mathaccent 94 p} + m) = 0$] with the usual
relativistic normalization:
\be
\overline{v}_\alpha (p) \gamma^\mu v_\beta (p) = 2 p^\mu \delta_{\alpha\beta}
~~~ , ~~~ \overline{v}_\alpha (p) v_\beta (p) = -2m \delta_{\alpha\beta} ~.
\ee
In fact, one can easily derive the following expression for the $n$--th order
term in the perturbative expansion ($n \ge 1$):
\ba
\lefteqn{
[\overline{v}_\beta (p) \tilde{G}^{(tc)} (\overline{\psi}_{p,j}
\to \overline{\psi}_{p',i}|A) v_\alpha (p')]_{(n)}
= } \nonumber \\
& & = \displaystyle\int {d^4 q_1 \over (2\pi)^4} \ldots
\displaystyle\int {d^4 q_n \over (2\pi)^4}
~(2\pi)^4 \delta^{(4)} (q - q_1 - \ldots - q_n)
~\overline{N}^{(ferm)}_{\alpha\beta,~ij} (q_1, \ldots, q_n) \nonumber \\
& & \times {i \over (p + q_1)^2 - m^2 + i\varepsilon}
\ldots {i \over (p + q_1 + \ldots + q_{n-1})^2 - m^2 + i\varepsilon} ~,
\label{anti_fn}
\ea
where $\overline{N}^{(ferm)}_{\alpha\beta,~ij} (q_1, \ldots, q_n)$
is given by:
\ba
\lefteqn{
\overline{N}^{(ferm)}_{\alpha\beta,~ij} (q_1, \ldots ,q_n) \equiv }
\nonumber \\
& & \equiv \overline{v}_\beta (p)
\{ [ i g \gamma^{\mu_1} \tilde{A}_{\mu_1} (q_1)
- i \delta m ~(2\pi)^4 \delta^{(4)} (q_1) \cdot {\bf 1} ]
({\mathaccent 94 p} + {\mathaccent 94 q}_1 - m) \ldots
\nonumber \\
& & ({\mathaccent 94 p} + {\mathaccent 94 q}_1 + \ldots
+ {\mathaccent 94 q}_{n-1} - m)
[ i g \gamma^{\mu_n} \tilde{A}_{\mu_n} (q_n)
- i \delta m ~(2\pi)^4 \delta^{(4)} (q_n) \cdot {\bf 1} ] \}_{ji}
v_\alpha (p') ~.
\ea
Proceeding as for the derivation of Eq. (\ref{f_num_fin}) and making use of
Eq. (\ref{cstar}), we find:
\ba
\lefteqn{
\overline{N}^{(ferm)}_{\alpha\beta,~ij} (q_1, \ldots ,q_n) \simeq }
\nonumber \\
& & \simeq \delta_{\alpha\beta} \cdot
\{ [ig 2p^{\mu_1} \tilde{A}_{\mu_1} (q_1)] \ldots
[ig 2p^{\mu_n} \tilde{A}_{\mu_n} (q_n)] \}_{ji} \nonumber \\
& & = \delta_{\alpha\beta} \cdot \{ [ig 2p^{\mu_n} \tilde{A}^*_{\mu_n} (q_n)]
\ldots [ig 2p^{\mu_1} \tilde{A}^*_{\mu_1} (q_1)] \}_{ij} ~.
\label{anti_num}
\ea
Apart from the {\it delta} function in front, this is exactly the term we would
have expected at the numerator of the $n$--th order term in the perturbative
expansion of the truncated--connected scalar antiquark propagator
$\tilde{S}^{(tc)} (\phi^\dagger_{p,j} \to \phi^\dagger_{p',i}|A)$ 
in the high--energy limit.
From Eqs. (\ref{anti_fn}) and (\ref{anti_num}) we derive
\be
[\overline{v}_\beta (p) \tilde{G}^{(tc)} (\overline{\psi}_{p,j}
\to \overline{\psi}_{p',i}|A) v_\alpha (p')]_{(n)} \simeq \delta_{\alpha\beta}
\cdot [\tilde{S}^{(tc)} (\phi^\dagger_{p,j} \to \phi^\dagger_{p',i}|A)]_{(n)}
~,
\ee
for every order $n$ in the perturbative expansion, so proving Eq.
(\ref{anti_f_fin}). As for the scalar case, going from quarks to antiquarks
corresponds just to the change from the fundamental representation $T_a$ of
$SU(N_c)$ to the complex conjugate representation $T'_a = -T^*_a$.

\newsection{The scattering of gluons}

\noindent
By using the same techniques developed in the previous sections, we shall
evaluate the ``truncated--connected gluon propagator'' in a given external
gluon field $A^\nu_b$, in the eikonal approximation. We call this quantity
``$\tilde{D}^{(tc)}_{\mu'\mu,~a'a} (k,k'|A)$'', whose $n$--th order
perturbative term is defined schematically in Fig. 1b, where the external
legs are supposed to be truncated on--shell ($k^2,k'^2 \to 0$).
More precisely, we shall evaluate the following quantity:
\be
\varepsilon^{\mu'*}_{(\lambda')} (k') \tilde{D}^{(tc)}_{\mu'\mu,~a'a} (k,k'|A)
\varepsilon^{\mu}_{(\lambda)} (k) ~,
\label{g_matrix}
\ee
where $\varepsilon^{\mu}_{(\lambda)} (k)$ are the polarization four--vectors
($\lambda,\lambda' \in \{ 1,2 \}$):
\be
\varepsilon_{(\lambda)} (k) \cdot \varepsilon^*_{(\lambda')} (k)
= -\delta_{\lambda\lambda'} ~~~ , ~~~ k \cdot
\varepsilon_{(\lambda)} (k) \vert_{k^2 = 0}  = 0 ~.
\label{polar}
\ee
This quantity should describe (under certain approximations that will be
discussed below and also in the next section) the scattering matrix element
of a gluon in a given external gluon field $A^\nu_b$:
\be
g(k,a,\lambda) \to g(k',a',\lambda') ~;
\ee
$a,a' \in \{ 1,\ldots, N_c^2 - 1 \}$ are colour indices and
$\lambda,\lambda' \in \{ 1,2 \}$ are spin indices.
In the eikonal approximation the dominant interaction between the incident
gluon and the external gluon field is represented by the
three--gluon vertex, which is linear in the four--momentum of the gluon
(while the four--gluon vertex is not dependent on the momentum).
Its Feynman rule is given by:
\ba
\lefteqn{
V^{a_1 a_2 a_3}_{\mu_1 \mu_2 \mu_3} (k_1,k_2,k_3) = } \nonumber \\
& & = g f^{a_1 a_2 a_3} [ g_{\mu_1 \mu_2} (-k_1 + k_2)_{\mu_3}
+ g_{\mu_2 \mu_3} (-k_2 - k_3)_{\mu_1}
+ g_{\mu_3 \mu_1} (k_3 + k_1)_{\mu_2} ] ~,
\label{3g_vertex}
\ea
where the four--momenta $k_1$ and $k_2$ are taken to be flowing into the
vertex, while the four--momentum $k_3$ is taken to be flowing out from the
vertex. The explicit expression of the $n$--th order perturbative term of the
quantity (\ref{g_matrix}), which is schematically defined in Fig. 1b, is
given by:
\ba
\lefteqn{
\left[ \varepsilon^{\mu'*}_{(\lambda')} (k') \tilde{D}^{(tc)}_{\mu'\mu,~a'a}
(k,k'|A) \varepsilon^{\mu}_{(\lambda)} (k) \right]_{(n)} \simeq } \nonumber \\
& & \simeq \displaystyle\int {d^4 q_1 \over (2\pi)^4} \ldots
\displaystyle\int {d^4 q_n \over (2\pi)^4}
~(2\pi)^4 \delta^{(4)} (q - q_1 - \ldots - q_n)
~N^{(gluon)}_{\lambda'\lambda,~a'a}
(q_1, \ldots, q_n) \nonumber \\
& & \times {i \over (k + q_1 + \ldots + q_{n-1})^2 + i\varepsilon}
\ldots {i \over (k + q_1)^2 + i\varepsilon} ~,
\label{gn_1}
\ea
where $q \equiv k' - k$ is the transferred momentum and
$N^{(gluon)}_{\lambda'\lambda,~a'a} (q_1, \ldots, q_n)$ is given by:
\ba
\lefteqn{
N^{(gluon)}_{\lambda'\lambda,~a'a} (q_1, \ldots, q_n) \equiv } \nonumber \\
& & \equiv \varepsilon^{\mu}_{(\lambda)} (k)
V^{a b_1 c_1}_{\mu \nu_1 \rho_1} (k,q_1,k+q_1) \tilde{A}^{\nu_1}_{b_1} (q_1)
g^{\rho_1 \rho_2} \delta_{c_1 c_2}
V^{c_2 b_2 c_3}_{\rho_2 \nu_2 \rho_3} (k+q_1,q_2,k+q_1+q_2) \times \ldots
\nonumber \\
& & \ldots \times
V^{c_{2n-2} b_n a'}_{\rho_{2n-2} \nu_n \mu'} (k+q_1+\ldots+q_{n-1},q_n,k')
\tilde{A}^{\nu_n}_{b_n} (q_n) \varepsilon^{\mu'*}_{(\lambda')} (k') ~.
\label{g_num}
\ea
It is not difficult to verify that, in the eikonal approximation, using
the properties (\ref{polar}) for the polarization four--vectors, the quantity
(\ref{g_num}) simplifies as follows:
\be
N^{(gluon)}_{\lambda'\lambda,~a'a} (q_1, \ldots, q_n) \simeq
\delta_{\lambda'\lambda} \cdot \{ [-ig 2k^{\mu_n} \tilde{\cal A}_{\mu_n} (q_n)]
\ldots [-ig 2k^{\mu_1} \tilde{\cal A}_{\mu_1} (q_1)] \}_{a'a} ~,
\label{g_num_fin}
\ee
where we have used the notation:
\be
{\cal A}_\mu \equiv A^b_\mu T^b_{(adj)} ~,
\ee
$T^b_{(adj)}$ being the $N_c^2-1$ matrices of the $SU(N_c)$ Lie algebra
in the adjoint representation:
\be
\left( T^a_{(adj)} \right)_{bc} = -i f^{abc} ~.
\ee
The expression (\ref{gn_1}), with the result (\ref{g_num_fin}), is perfectly
analogous to the corresponding expression appearing in Eq. (\ref{sn_1}) for
the case of the scalar quark. By proceeding exactly as in Sect. 2, we can thus
further approximate the above--written Eq. (\ref{gn_1}) as follows:
\ba
\lefteqn{
\left[ \varepsilon^{\mu'*}_{(\lambda')} (k') \tilde{D}^{(tc)}_{\mu'\mu,~a'a}
(k,k'|A) \varepsilon^{\mu}_{(\lambda)} (k) \right]_{(n)} \simeq } \nonumber \\
& & \simeq \delta_{\lambda'\lambda} \cdot 2E \displaystyle\int [d^3 b]
~e^{iq b} \displaystyle\int d\tau_1 \ldots \displaystyle\int d\tau_n
~\theta (\tau_n - \tau_{n-1}) \ldots \theta (\tau_2 - \tau_1)
\nonumber \\
& & \times \{ \left[ -ig k^{\mu_n} {\cal A}_{\mu_n} (b + k\tau_n) \right]
\ldots \left[ -ig k^{\mu_1} {\cal A}_{\mu_1} (b + k\tau_1) \right] \}_{a'a} ~.
\ea
Therefore, summing all orders ($n \ge 1$), we finally obtain:
\ba
\lefteqn{
\varepsilon^{\mu'*}_{(\lambda')} (k') \tilde{D}^{(tc)}_{\mu'\mu,~a'a} (k,k'|A)
\varepsilon^{\mu}_{(\lambda)} (k) \simeq } \nonumber \\
& & \simeq \delta_{\lambda'\lambda} \cdot 2E \displaystyle\int [d^3 b]
~e^{iqb} [{\cal V}_k (b) - {\bf 1}]_{a'a} ~,
\label{g_fin}
\ea
where ${\cal V}_k (b)$ is the Wilson string along the path $x(\tau) =
b + k\tau$ ($\tau \in [-\infty,+\infty]$), in the adjoint representation,
defined as:
\ba
\lefteqn{
{\cal V}_k (b) \equiv T \exp \left( -ig \displaystyle\int_{-\infty}^{+\infty}
{\cal A}_\mu (b + k\tau) k^\mu d\tau \right) \equiv } \nonumber \\
& & \equiv \displaystyle\sum_{n = 0}^{\infty}
\displaystyle\int d\tau_1 \ldots \displaystyle\int d\tau_n
~\theta (\tau_n - \tau_{n-1}) \ldots \theta (\tau_2 - \tau_1)
\nonumber \\
& & \times \left[ -ig k^{\mu_n} {\cal A}_{\mu_n} (b + p\tau_n) \right] \ldots
\left[ -ig k^{\mu_1} {\cal A}_{\mu_1} (b + k\tau_1) \right] ~.
\label{v_def}
\ea
Eq. (\ref{g_fin}) gives the expression for the scattering matrix element of a
gluon in a given external gluon field, in the eikonal approximation.

\newsection{Parton--parton scattering amplitudes}

\noindent
In the previous sections, we have derived expressions for the
truncated--connected quark (antiquark) propagator in a given external gluon
field $A^\mu$, by a direct resummation of perturbation theory in the limit of
very high energy and small transferred momentum. The procedure has been also
extended to include the case in which the scattering parton is a gluon:
additional approximations are necessary and they will be discussed below.

The truncated--connected propagators in an external gluon field are the basic
ingredients to derive high--energy parton-parton scattering amplitudes,
using the LSZ reduction formulae and a functional integral approach.
This was done in Ref. \cite{Meggiolaro96} and it will
be quickly reviewed here for the convenience of the reader.
Let us consider, for example, the elastic scattering process of two scalar
quarks with initial four--momenta $p_1$ and $p_2$ and final four--momenta
$p'_1$ and $p'_2$:
\be
\phi_j (p_1) + \phi_l (p_2) \rightarrow \phi_i (p'_1) + \phi_k (p'_2) ~.
\ee
[$i,j,k,l \in \{ 1, \ldots, N_c \}$ are colour indices.]
In the center--of--mass reference system (c.m.s.), taking the initial
trajectories of the two quarks along the $x^1$--axis, the four--momenta
$p_1$, $p_2$, $p'_1$ and $p'_2$ are given, in the limit of ``soft''
high--energy scattering, $s = (p_1 + p_2)^2 \to \infty$ and
$t = (p_1 - p'_1)^2 \ll s$, by:
\be
p_1 \simeq p'_1 \simeq (E,E,\vec{0}_t) ~~~ , ~~~
p_2 \simeq p'_2 \simeq (E,-E,\vec{0}_t) ~.
\ee
Using the LSZ reduction formulae and a functional integral approach, one
finds the following expression for the scattering matrix element [with the
plane wave functions normalized as: $\phi_p (x) = \exp(-ipx)$]:
\ba
\lefteqn{
\langle \phi_i (p'_1) \phi_k (p'_2) | ( S - {\bf 1} ) |
\phi_j (p_1) \phi_l (p_2) \rangle \simeq } \nonumber \\
& & \simeq {1 \over Z_W^2} \{ \langle \tilde{S}^{(tc)}_{ij} (p_1,p'_1|A)
\tilde{S}^{(tc)}_{kl} (p_2,p'_2|A) \rangle_A +
\langle \tilde{S}^{(tc)}_{kj} (p_1,p'_2|A)
\tilde{S}^{(tc)}_{il} (p_2,p'_1|A) \rangle_A \} ~,
\label{s_scatt}
\ea
where $Z_W$ is the residue at the pole (i.e., for $p^2 \to m^2$) of the
unrenormalized quark propagator [see Eq. (\ref{s_mass})].
The expectation value $\langle O (A) \rangle_A$ of an arbitrary functional
$O (A)$ of the gluon field $A^\mu$ is defined as:
\be
\langle O (A) \rangle_A \equiv {1 \over Z_{QCD}} \displaystyle\int [dA] ~O (A)
\exp \left[ -{i \over 4} \displaystyle\int d^4 x ~F_a^{\mu\nu} F_{a \mu\nu}
\right] \{ \det [ D^\mu D_\mu + m^2 ] \}^{-1} ~,
\label{s_exp}
\ee
where $Z_{QCD}$ is the partition function for scalar QCD:
\ba
\lefteqn{
Z_{QCD} \equiv \displaystyle\int [dA] [d\phi] [d\phi^\dagger]
\exp \left[ i \int d^4 x ~L (\phi ,\phi^\dagger ,A) \right] }
\nonumber \\
& & = \displaystyle\int [dA]
\exp \left[ -{i \over 4} \displaystyle\int d^4 x ~F_a^{\mu\nu} F_{a \mu\nu}
\right] \{ \det [ D^\mu D_\mu + m^2 ] \}^{-1} ~.
\label{s_part}
\ea
The determinant in Eqs. (\ref{s_exp}) and (\ref{s_part}) comes from the
integration over the scalar degrees of freedom.
In fact, the Lagrangian $L (\phi ,\phi^\dagger, A)$
is bilinear in the scalar fields $\phi$ and $\phi^\dagger$:
\be
\displaystyle\int d^4 x \left[ (D^\mu \phi)^\dagger D_\mu \phi - m^2
\phi^\dagger \phi \right] =
- \displaystyle\int d^4 x ~\phi^\dagger \left[ D^\mu D_\mu + m^2 \right]
\phi ~.
\ee
Therefore, the functional integral over the scalar fields $\phi$ and
$\phi^\dagger$ is an ordinary Gaussian integral and can be
performed in the standard way, to give (apart from an irrelevant constant)
$\{ \det [ D^\mu D_\mu + m^2 ] \}^{-1}$.

The first expectation value in Eq. (\ref{s_scatt}) corresponds to the
$t$--channel scattering of the two particles, with squared transferred
momentum equal to: $(p_1 - p'_1)^2 \equiv t \ll s$.
The second expectation value corresponds instead to the $u$--channel
scattering of the two quarks. In other words, the squared transferred
momentum, flowing from one quark to the other, is equal to:
\be
(p_1 - p'_2)^2 \equiv u = 4 m^2 -s -t \simeq -s ~,
\ee
in the limit $s \to \infty$ with $t,~m^2 \ll s$.
In this high--energy limit the contribution coming from the second expectation
value in Eq. (\ref{s_scatt}) is smaller by at least a factor of $s$, when
compared with the first expectation value, and hence is negligible.
[One can be easily convinced of this by considering the Feynman diagrams of
the process in the perturbation theory: the diagrams which correspond to the
second piece in Eq. (\ref{s_scatt}) have intermediate gluons carrying a big
squared transferred momentum $u \simeq -s$, so that their propagators
suppress the corresponding amplitude.]
Therefore, in our limit:
\be
\langle \phi_i (p'_1) \phi_k (p'_2) | ( S - {\bf 1} ) |
\phi_j (p_1) \phi_l (p_2) \rangle \simeq
{1 \over Z_W^2} \langle \tilde{S}^{(tc)}_{ij} (p_1,p'_1|A)
\tilde{S}^{(tc)}_{kl} (p_2,p'_2|A) \rangle_A ~,
\ee
where for the truncated--connected propagator $\tilde{S}^{(tc)} (p,p'|A)$ we
can use the expression (\ref{s_fin}) derived in Sect. 2 in the eikonal limit.
Also the quantity $Z_W$ can be evaluated in the same approximation: the
result and the details of the calculations are reported in the Appendix.
If one defines the diffusion amplitude $T_{fi} = < f | T | i >$ by
\ba
\lefteqn{
\langle \phi_i (p'_1) \phi_k (p'_2) | ( S - {\bf 1} ) |
\phi_j (p_1) \phi_l (p_2) \rangle = } \nonumber \\
& & = i (2\pi)^4 \delta^{(4)} (P_{fin} - P_{in})
~\langle \phi_i (p'_1) \phi_k (p'_2) | T | \phi_j (p_1) \phi_l (p_2) \rangle ~,
\ea
where $P_{in} = p_1 + p_2$ is the initial total four--momentum and
$P_{fin} = p'_1 + p'_2$ is the final total four--momentum, the following
results is obtained at the end \cite{Meggiolaro96}:
\ba
\lefteqn{
\langle \phi_i (p'_1) \phi_k (p'_2) | T | \phi_j (p_1) \phi_l (p_2) \rangle
\simeq } \nonumber \\
& & \simeq -{i \over Z_W^2} 2s
\displaystyle\int d^2 \vec{z}_\perp ~e^{i \vec{q}_\perp \cdot \vec{z}_\perp}
\langle [ W_{p_1} (z_t) - {\bf 1} ]_{ij} [ W_{p_2} (0) - {\bf 1} ]_{kl}
\rangle_A ~,
\label{s_scatt_fin}
\ea
where $q = p_1 - p'_1 \simeq (0,0,\vec{q}_\perp)$, with
$t = q^2 = -\vec{q}_\perp^2$, is the transferred four--momentum, and
$z_t = (0,0,\vec{z}_\perp)$ is the distance between the two trajectories
in the transverse plane.

One can proceed exactly in the same way when considering the elastic scattering
process of two ``real'' (i.e., spin--1/2) quarks with initial four--momenta
$p_1$, $p_2$ and final four--momenta $p'_1$, $p'_2$, in the limit of ``soft''
high--energy scattering:
\be
\psi_{j\beta} (p_1) + \psi_{l\delta} (p_2) \rightarrow \psi_{i\alpha} (p'_1) +
\psi_{k\gamma} (p'_2) ~.
\ee
The following expression for the scattering matrix element is derived using
the LSZ reduction formulae and a functional integral approach [with the
plane wave functions normalized as: $\psi_p (x) = u(p) \exp(-ipx)$, with
$\overline{u}_\alpha (p) u_\beta (p) = 2m \delta_{\alpha\beta}$]:
\ba
\lefteqn{
\langle \psi_{i\alpha}(p'_1) \psi_{k\gamma}(p'_2) | ( S - {\bf 1} ) |
\psi_{j\beta}(p_1) \psi_{l\delta}(p_2) \rangle \simeq }
\nonumber \\
& & \simeq {1 \over Z_W^2} \langle
\overline{u}_\alpha (p'_1) \tilde{G}^{(tc)}_{ij} (p_1, p'_1|A)
u_\beta (p_1) \cdot \overline{u}_\gamma (p'_2)
\tilde{G}^{(tc)}_{kl} (p_2, p'_2|A) u_\delta (p_2) \rangle_A ~.
\ea
[$i,j,k,l \in \{ 1, \ldots, N_c \}$ are colour indices, while $\alpha,\beta,
\gamma,\delta \in \{ 1,2 \}$ are spin indices.]

The expectation value $\langle O (A) \rangle_A$ of an arbitrary functional
$O (A)$ of the gluon field $A^\mu$ is now defined as:
\be
\langle O (A) \rangle_A \equiv {1 \over Z_{QCD}} \displaystyle\int [dA] ~O (A)
\exp \left[ -{i \over 4} \displaystyle\int d^4 x ~F_a^{\mu\nu} F_{a \mu\nu}
\right] \det [ i\gamma^\mu D_\mu - m ] ~,
\label{f_exp}
\ee
where $Z_{QCD}$ is the partition function for fermion QCD:
\ba
\lefteqn{
Z_{QCD} \equiv \displaystyle\int [dA] [d\psi] [d\psi^\dagger]
\exp \left[ i \int d^4 x ~L (\psi ,\psi^\dagger ,A) \right] }
\nonumber \\
& & = \displaystyle\int [dA]
\exp \left[ -{i \over 4} \displaystyle\int d^4 x ~F_a^{\mu\nu} F_{a \mu\nu}
\right] \det [ i\gamma^\mu D_\mu - m ] ~.
\label{f_part}
\ea
The determinant in Eqs. (\ref{f_exp}) and (\ref{f_part}) comes from the
integration over the fermion degrees of freedom.

If one uses the result (\ref{f_fin}) for $\overline{u}_\alpha (p')
\tilde{G}^{(tc)}_{ij} (p,p'|A) u_\beta (p)$, one finally finds the
following expression for the high--energy quark--quark elastic scattering
amplitude in (fermion) QCD \cite{Nachtmann91,Nachtmann97,Meggiolaro96}:
\ba
\lefteqn{
\langle \psi_{i\alpha}(p'_1) \psi_{k\gamma}(p'_2) | T |
\psi_{j\beta}(p_1) \psi_{l\delta}(p_2) \rangle \simeq } \nonumber \\
& & \simeq -{i \over Z_W^2} \cdot \delta_{\alpha\beta} \delta_{\gamma\delta}
\cdot 2s
\displaystyle\int d^2 \vec{z}_\perp ~e^{i \vec{q}_\perp \cdot \vec{z}_\perp}
\langle [ W_{p_1} (z_t) - {\bf 1} ]_{ij} [ W_{p_2} (0) - {\bf 1} ]_{kl}
\rangle_A ~.
\label{f_scatt_fin}
\ea
The notation is the same as for Eq. (\ref{s_scatt_fin}). In a perfectly
analogous way, one can also derive the high--energy scattering amplitude
for an elastic process involving two partons, which can be quarks, antiquarks
or gluons. One simply has to insert in the matrix element, for each of the
two partons involved, the corresponding quantity among (\ref{s_fin}),
(\ref{anti_s_fin}), (\ref{f_fin}), (\ref{anti_f_fin}) and (\ref{g_fin}),
which, as we have said before, describe the scattering amplitude of the
parton in a given external gluon field. So, for example, the amplitude
for the quark--gluon scattering
\be
\psi_{j\beta} (p_1) + g_{a\lambda} (p_2) \rightarrow \psi_{i\alpha} (p'_1) +
g_{a'\lambda'} (p'_2)
\ee
has the following expression in the high--energy limit \cite{Nachtmann97}:
\ba
\lefteqn{
\langle \psi_{i\alpha}(p'_1) g_{a'\lambda'}(p'_2) | T |
\psi_{j\beta}(p_1) g_{a\lambda}(p_2) \rangle \simeq } \nonumber \\
& & \simeq {1 \over Z_W Z_{\cal V}} \langle
\overline{u}_\alpha (p'_1) \tilde{G}^{(tc)}_{ij} (p_1, p'_1|A)
u_\beta (p_1) \cdot \varepsilon^{\mu'*}_{(\lambda')} (p'_2)
\tilde{D}^{(tc)}_{\mu'\mu,~a'a} (p_2,p'_2|A)
\varepsilon^{\mu}_{(\lambda)} (p_2) \rangle \nonumber \\
& & \simeq -{i \over Z_W Z_{\cal V}} \cdot \delta_{\alpha\beta}
\delta_{\lambda'\lambda} \cdot 2s
\displaystyle\int d^2 \vec{z}_\perp ~e^{i \vec{q}_\perp \cdot \vec{z}_\perp}
\langle [ W_{p_1} (z_t) - {\bf 1} ]_{ij} [ {\cal V}_{p_2} (0) - {\bf 1} ]_{a'a}
\rangle_A ~.
\label{fg_scatt_fin}
\ea
(The renormalization constant $Z_{\cal V}$ is defined and evaluated in
the Appendix.)
Let us observe at this point that, differently from their quark counterparts
(which really come from the integration over the quark degrees of freedom
in the functional integral), the gluon matrix element
$\varepsilon^{\mu'*}_{(\lambda')} (k') \tilde{D}^{(tc)}_{\mu'\mu,~a'a} (k,k'|A)
\varepsilon^{\mu}_{(\lambda)} (k)$ can only be defined as the proper functional
of $A^\mu$ which, when inserted in the functional average $\langle \ldots
\rangle_A$ (together with another parton matrix element) reproduces the
corresponding gluon--parton scattering amplitude in the high--energy limit.
We have derived the expression (\ref{g_fin}) for the gluon matrix element by
resumming all diagrams of the type reported in Fig. 1b in the eikonal limit.
Therefore, e.g., Compton--like diagrams (and their perturbative corrections)
are not included in Eq. (\ref{fg_scatt_fin}): indeed, one can easily convince
oneself (see, for example, Ref. \cite{Cheng-Wu-book}) that these diagrams
(which are present also in the Abelian case) are of order ${\cal O} (s^0)$
in the high--energy limit $s \to \infty$; while, for example, the
${\cal O} (g^2)$ diagram described by Eq. (\ref{fg_scatt_fin}) (i.e., the
one obtained taking the first term in the perturbative expansion of the gluon
matrix element and of the quark matrix element) is of order ${\cal O} (s)$.
However, the expression (\ref{fg_scatt_fin}), by its own construction, is not
able to reproduce all those diagrams where the scattering gluon converts in a
quark--antiquark pair during the diffusion process. Without resorting to
large--$N_c$ approximations, where these diagrams are of course sub--leading
corrections, a reason for not including these diagrams in the high--energy
scattering amplitude can come if we first study ``soft'' high--energy
parton--parton scattering in the ``femto universe'', in the spirit of Refs.
\cite{Nachtmann91,Nachtmann97}. In other words, we first consider the
scattering of the partons over the finite time interval
$-t_0/2 \le t \le t_0/2$ ($t = 0$ being the nominal collision time of the
hadrons $h_1 + h_2 \to h_1 + h_2$ in the the c.m.s.) of length
$t_0 \approx 2$ fm: $t_0/2$ is the time when, in an inelastic collision,
the first produced hadrons appear.
(The estimate $t_0 \approx 2$ fm is discussed in Ref. \cite{Nachtmann91}.)
As discussed in Refs. \cite{Nachtmann91,Nachtmann97}, one can assume that
over that time interval: a) parton annihilation and production processes
can be neglected (i.e., the parton state of the hadrons does not change
qualitatively in this time); b) partons travel in essence on straight
lightlike world lines and they undergo ``soft'' elastic scattering.
This was the strategy adopted in Ref. \cite{Nachtmann91} in order to study
parton--parton ``soft'' high--energy scattering in QCD. Of course, free
quarks and gluons do not exist in (zero--temperature) QCD, but, assuming
that $t_0 \approx 2$ fm is nearly infinitely long on the scale of the
``femto universe'', one can use the standard LSZ reduction formulae to
relate the partonic $S$--matrix element to an integral over the four--point
function of the quark/gluon fields.
After having solved the problem of parton--parton scattering, one has to fold
the partonic $S$--matrix with the hadronic wave functions of the appropriate
resolution to get the hadronic $S$--matrix elements.
We want to stress that the result (\ref{f_scatt_fin}) for quark--quark
scattering at high energies was derived in Ref. \cite{Nachtmann91} under
the crucial assumption that only gluon modes up to a fixed frequency
contribute in the functional integral $\langle \ldots \rangle_A$.
It was argued at length in Ref. \cite{Nachtmann91} that this should be a
valid approximation for the scattering of partons over the time interval
$(-t_0/2,t_0/2)$, since, for these scattering processes, the relevant scale
for the frequency of the exchanged quanta is given by $a^{-1}$, indipendent
of $s$, where $a$ is the correlation length of the (gauge--invariant)
two--point function of the gluon field--strength tensor. In the hypothetical
scattering amplitude for ``real'' quarks, all the splitting processes with
long time scales will play an important role and therefore must be included.

\newsection{Conclusions and outlook}

\noindent
In this paper we have derived the same results already found in Refs.
\cite{Nachtmann91,Nachtmann97,Meggiolaro96} in a different and even more
immediate way, i.e., by a direct resummation of perturbation theory in the
limit of very high energy and small transferred momentum.
The approximations used in the previous sections in order to derive the
eikonal propagators (and, as a consequence, the parton--parton scattering
amplitudes) in the above--mentioned limit are exactly the same adopted in
Ref. \cite{Cheng-Wu-book} in order to derive the high--energy asymptotic
expressions for the elastic parton--parton scattering amplitudes at a given
order in perturbation theory, using standard Feynman--diagrams techniques.
(These approximations essentially consist in neglecting the recoil of the
scattering partons in the diffusion process.)
Therefore, in this sense, we can claim that we have given a proof that the
nonperturbative expressions (\ref{s_scatt_fin}), (\ref{f_scatt_fin}), etc., 
for the high--energy elastic scattering amplitude are a resummation of the
corresponding perturbative results found in Ref. \cite{Cheng-Wu-book}.
(In a previous paper \cite{Meggiolaro97}, we gave a direct proof that this is
true up to the fourth order in the expansion in the coupling constant.)

We remind the reader that the $s$ dependence of the scattering amplitude
is not all contained in the kinematical factor $2s$ in front of the
integral in Eqs. (\ref{s_scatt_fin}), (\ref{f_scatt_fin}), etc.~.
In fact, as was first pointed out by Verlinde and Verlinde in \cite{Verlinde},
it is a singular limit to take the Wilson lines in (\ref{s_scatt_fin}),
(\ref{f_scatt_fin}), etc., exactly lightlike. It turns out that a
proper regularization of these ``infrared'' singularities (so called
because they essentially come from the limit $m \to 0$, $m$ being the
quark mass) gives rise to a $\log s$ dependence of the amplitude,
as obtained by ordinary perturbation theory \cite{Cheng-Wu-book,Lipatov}
and as confirmed by the experiments on hadron--hadron scattering
processes. In practice, the regularization procedure consists in letting
each Wilson line have a small timelike component (so that they coincide
with the classical trajectories for quarks with a finite mass $m$) and
letting them end after some finite proper time $\pm T$, i.e., after some time
$|t| \sim T \sqrt{s}/2m$ in the c.m.s. (so that, when one takes the limit
$s \to \infty$, the trajectories of the Wilson lines will again become
infinitely long and lightlike). We refer the reader to Refs. \cite{Verlinde}
and \cite{Meggiolaro97,Meggiolaro98,Meggiolaro-proc} for a detailed
discussion about this point.

We want also to stress that the expressions (\ref{s_scatt_fin}),
(\ref{f_scatt_fin}), etc., for the scattering amplitudes are not limited by
our approximations to be {\it ``quenched''}: indeed, in the functional average
$\langle \ldots \rangle_A$ also the determinant of the quark matrix is
included and it gives rise to Feynman diagrams with dynamical quark loops
in the perturbative expansion.
So, for example, the so--called {\it ``tower diagrams''} for quark--quark
elastic scattering, which give amplitudes as large as $s (\log s)^n$, where
$n$ is the number of quark loops joined together vertically, with each loop
having four vertices (see, e.g., Ref. \cite{Cheng-Wu-book} and references
therein), are expected to be reproduced in the perturbative expansion of
Eqs. (\ref{s_scatt_fin}), (\ref{f_scatt_fin}), etc., provided you integrate
from $t = -\infty$ to $t = +\infty$, i.e., you take infinitely long Wilson
lines, since the time associated with the hard fluctuations in these eikonal
processes is very large.
[If we truncate the Wilson lines from $t = -t_0$ to $t = +t_0$
(in the c.m.s. of the reaction), with $t_0$ fixed as $s \to \infty$,
we can loose in our description all those processes in which there is
production of hard partons, whose fluctuation time is larger than $t_0$
(in the asymptotic limit $s \to \infty$). These processes must then be
included in the wave functions of the fermions at $t = \pm t_0$.]

As we have said before, free asymptotic states of quarks and gluons do not
exist in (zero--temperature) QCD. Therefore, in this case, the correct 
procedure should be the one suggested in Refs. \cite{Nachtmann91,Nachtmann97},
where Eqs. (\ref{s_scatt_fin}), (\ref{f_scatt_fin}), etc., are considered to
be the partonic $S$--matrix elements in the ``femto universe'' (with a fixed
frequency cutoff for the gluon modes in the functional average): they must be
folded with the hadronic wave functions of the appropriate resolution to get
the hadronic $S$--matrix elements.

Viceversa, in QED one can have the free--electron states as IN and OUT
asymptotic states: there is no complication due to confinement and a
free--electron state is a well--defined asymptotic state of the theory.
Therefore, the eikonal formula (\ref{f_scatt_fin}) in QED, with infinitely
long Wilson lines, can be really considered as the asymptotic expression for
the fermion--fermion elastic scattering amplitude in the high--energy limit
and it should reproduce all perturbative results evaluated by Cheng and Wu
in Ref. \cite{Cheng-Wu-book}.
We want to stress that the eikonal formula (\ref{f_scatt_fin}) is not
identical to what, in the literature, is usually called the ``eikonal
amplitude'' of the high--energy scattering in QED (see Refs.
\cite{Cheng-Wu,Abarbanel-Itzykson,Jackiw}): this last was obtained
in the so--called {\it ``quenched''} approximation, where vacuum polarization
effects, arising from the presence of dynamical fermion loops, are
neglected. It was proved in Ref. \cite{Meggiolaro96} (see also Refs.
\cite{Korchemsky,Arefeva94}) that, when evaluating Eq. (\ref{f_scatt_fin})
in the {\it quenched} approximation, one correctly reproduces the eikonal
result of Refs. \cite{Cheng-Wu,Abarbanel-Itzykson,Jackiw}. However,
as we have said before, the amplitude (\ref{f_scatt_fin}) is not
limited, in general, to be {\it quenched} and so it is expected to be
more correct and to contain more information than the eikonal result
of Refs. \cite{Cheng-Wu,Abarbanel-Itzykson,Jackiw}).

Once we have found the nonperturbative expressions (\ref{s_scatt_fin}),
(\ref{f_scatt_fin}), etc., for the high--energy scattering amplitudes,
the natural question which arises is: How can we evaluate them directly?
The answer to this question is highly nontrivial and it is also strictly
connected with the renormalization properties of Wilson--line operators
\cite{Arefeva80,Korchemsky}.
Some nonperturbative approaches for the calculation of (\ref{f_scatt_fin})
were proposed in Refs. \cite{Arefeva94} and \cite{Dosch}. In particular,
in Refs.  \cite{Dosch,Berger} a nonperturbative numerical estimate of
high--energy hadron--hadron scattering amplitudes was obtained, using Eq.
(\ref{f_scatt_fin}) as a basic ingredient, in the framework of the so--called
``stochastic vacuum model''.
For an alternative approach to the problem, we refer the reader to Refs.
\cite{Meggiolaro97,Meggiolaro98,Meggiolaro-proc}, where interesting analytic
properties of the high--energy scattering amplitude were derived, going from
Minkowskian to Euclidean space--time, so opening the possibility of studying
the high--energy scattering amplitude using the Euclidean formulation of the
theory, e.g., on the lattice. (See also Ref. \cite{HMN}, where a similar
analytic continuation from Minkowskian to Euclidean theory was proposed
to study the small-$x_{\rm Bj}$ behaviour of the structure functions of
deep inelastic lepton--nucleon scattering.)
The analytic continuation proposed in Refs.
\cite{Meggiolaro97,Meggiolaro98,Meggiolaro-proc} has been recently adopted
in Ref. \cite{Janik-Peschanski}, in order to study the high--energy
scattering in ${\cal N} = 4$ supersymmetric $SU(N_c)$ gauge theories
(in the strong coupling, large--$N_c$ limit) using the AdS/CFT correspondence,
and also in Ref. \cite{Shuryak-Zahed}, in order to investigate
instanton--induced effects in QCD high--energy scattering.
In our opinion, a considerable progress could be achieved by a direct
investigation of these problems on the lattice in the near future.

\bigskip
\noindent {\bf Acknowledgements}
\smallskip
 
I would like to thank Prof. Otto Nachtmann for his useful suggestions
and comments and also for having encouraged me to write this paper. 

\vfill\eject

\renewcommand{\thesection}{}
\renewcommand{\thesubsection}{A.\arabic{subsection} }
 
\pagebreak[3]
\setcounter{section}{1}
\setcounter{equation}{0}
\setcounter{subsection}{0}
\setcounter{footnote}{0}

\begin{flushleft}
{\large\bf \thesection Appendix: The residue at the pole of the unrenormalized
eikonal propagators}
\end{flushleft}

\renewcommand{\thesection}{A}

\setcounter{subsection}{1}
\begin{flushleft}
{\large\bf \thesubsection The scalar propagator}
\end{flushleft}

\noindent
Let us consider now the unrenormalized full scalar propagator (not truncated!)
in the vicinity of the pole (i.e., for $p^2 \to m^2$). We start from the
scalar propagator $S_{ij} (x,y|A)$ in an external gluon field $A^\mu$
and then we take the functional average over the 
gluon field in order to get the full propagator (\ref{s_prop}):
\be
\langle S_{ij} (x,y|A) \rangle_A = \langle T [ \phi_i(x) \phi^\dagger_j(y) ]
\rangle = S_{ij} (x-y) ~.
\ee
Therefore, using the definition (\ref{fourier}), we find:
\be
\langle \tilde{S}_{ij} (p,p'|A) \rangle_A = (2\pi)^4 \delta^{(4)} (p' - p)
~\tilde{S}_{ij} (p) ~,
\label{as_prop}
\ee
where $\tilde{S}_{ij} (p)$ is the unrenormalized propagator in the momentum
space, defined by Eq. (\ref{s_tilde}).
We shall evaluate the quantity (\ref{as_prop}) starting
from the perturbative expansion of $\tilde{S}^{(tc)}_{ij} (p,p'|A)$ in the
eikonal approximation, that we have computed in Sect. 2.
At the 0--th order we find the following expression for
$\langle \tilde{S}_{ij} (p,p'|A) \rangle_A$:
\be
[ \langle \tilde{S}_{ij} (p,p'|A) \rangle_A ]_{(0)} =
(2\pi)^4 \delta^{(4)} (p' - p)
~{i \over p^2 - m^2 + i\varepsilon} ~.
\label{asn_0}
\ee
At the $n$--th order ($n \ge 1$) we find, using the corresponding expression
for $[\tilde{S}^{(tc)}_{ij} (p,p'|A)]_{(n)}$ that we have derived in
Sect. 2 [see Eq. (\ref{sn_2})]:
\ba
\lefteqn{
[ \langle \tilde{S}_{ij} (p,p'|A) \rangle_A ]_{(n)} \simeq } \nonumber \\
& & \simeq {i \over p'^2 - m^2 + i\varepsilon}
\displaystyle\int d^2 \vec{b}_{n\perp} \displaystyle\int d b_{n-}
\displaystyle\int d b_{n+} \ldots \displaystyle\int d b_{1+} ~e^{iq b_n}
\nonumber \\
& & \times {1 \over (2E)^{n-1}} \theta (b_{n+} - b_{(n-1)+}) \ldots
\theta (b_{2+} - b_{1+}) \nonumber \\
& & \times \langle \{ \left[ -ig p^{\mu_n} A_{\mu_n} (b_n) \right] \ldots
\left[ -ig p^{\mu_1} A_{\mu_1} (b_1) \right] \}_{ij} \rangle_A
\vert_{ b_{i-} = b_{n-} ~;
~\vec{b}_{i\perp} = \vec{b}_{n\perp} } \nonumber \\
& & \times {i \over p^2 - m^2 + i\varepsilon} ~,
\label{asn_1}
\ea
where $q = p' - p$.
By virtue of the invariance under the Poincar\`e group (in particular,
under space--time translations):
\be
\langle A_{\mu_n} (b_n) A_{\mu_{n-1}} (b_{n-1}) \ldots A_{\mu_1} (b_1)
\rangle_A = \langle A_{\mu_n} (0) A_{\mu_{n-1}} (b_{n-1} - b_n) \ldots
A_{\mu_1} (b_1 - b_n) \rangle_A ~.
\ee
Therefore, making the change of variables
\be
c_i = b_i - b_n, ~\forall i = 1, \ldots, n-1 ~; ~~ c_n = b_n ~,
\ee
the integration over $c_n$ gives $2 (2\pi)^4 \delta^{(4)} (q)$ and we can
write Eq. (\ref{asn_1}) as follows:
\ba
\lefteqn{
[ \langle \tilde{S}_{ij} (p,p'|A) \rangle_A ]_{(n)} \simeq } \nonumber \\
& & \simeq 2 (2\pi)^4 \delta^{(4)} (q) ~{i \over p'^2 - m^2 + i\varepsilon}
\displaystyle\int d c_{(n-1)+} \ldots \displaystyle\int d c_{1+}
\nonumber \\
& & \times {1 \over (2E)^{n-1}} \theta (-c_{(n-1)+})
\theta (c_{(n-1)+} - c_{(n-2)+}) \ldots
\theta (c_{2+} - c_{1+}) \nonumber \\
& & \times \langle \{ \left[ -ig p^{\mu_n} A_{\mu_n} (0) \right]
\left[ -ig p^{\mu_{n-1}} A_{\mu_{n-1}} (c_{n-1}) \right] \ldots
\left[ -ig p^{\mu_1} A_{\mu_1} (c_1) \right] \}_{ij} \rangle_A
\vert_{ c_{i-} = 0 ~; ~\vec{c}_{i\perp} = \vec{0}_\perp } \nonumber \\
& & \times {i \over p^2 - m^2 + i\varepsilon} ~.
\label{asn_2}
\ea
In the vicinity of the pole, i.e., for $p^2 \to m^2$, one has that:
$p'^2 - m^2 \simeq 2pq + q^2$.
Moreover, in the eikonal limit, $p_+ \simeq 2E$, $p_- \simeq 0$ and
$\vec{p}_\perp \simeq \vec{0}_\perp$, so that:
\be
{i \over p'^2 - m^2 + i\varepsilon} \simeq {i \over 2E}
{1 \over q_- + i\varepsilon} = {i \over 2E} {\cal P} {1 \over q_-}
+ {\pi \over 2E} \delta (q_-) ~,
\ee
where ``${\cal P}$'' stands for ``principal--part value''.
While $\delta^{(4)} (q) {\cal P} (1/q_-) = 0$, we can write, formally:
\be
2 \delta^{(4)} (q) {\pi \over 2E} \delta (q_-) =
\delta^{(4)} (q) {1 \over 2E} \displaystyle\int_{-\infty}^{+\infty}
dc_{n+} ~e^{ic_{n+} q_-} =
\delta^{(4)} (q) {1 \over 2E} \displaystyle\int_{-\infty}^{+\infty}
dc_{n+} ~.
\ee
(The origin of this infrared singularity is discussed in Sect. 6 and
can be ``cured'' by a proper regularization of the Wilson lines, as
suggested in Refs. \cite{Verlinde} and
\cite{Meggiolaro97,Meggiolaro98,Meggiolaro-proc}.)
Substituting in Eq. (\ref{asn_2}) and changing again the integration
variables from $c_i$ to $b_i = c_i + c_n, ~\forall i = 1, \ldots, n-1$
and $b_n = c_n$, we obtain:
\ba
\lefteqn{
[ \langle \tilde{S}_{ij} (p,p'|A) \rangle_A ]_{(n)} \simeq } \nonumber \\
& & \simeq (2\pi)^4 \delta^{(4)} (q) ~{i \over p^2 - m^2 + i\varepsilon}
\displaystyle\int d b_{n+} \ldots \displaystyle\int d b_{1+}
\nonumber \\
& & \times {1 \over (2E)^n} \theta (b_{n+} - b_{(n-1)+})
\ldots \theta (b_{2+} - b_{1+}) \nonumber \\
& & \times \langle \{ \left[ -ig p^{\mu_n} A_{\mu_n} (b_n) \right]
\ldots \left[ -ig p^{\mu_1} A_{\mu_1} (b_1) \right] \}_{ij} \rangle_A
\vert_{ b_{i-} = b_{n-} ~; ~\vec{b}_{i\perp} = \vec{b}_{n\perp} } ~.
\label{asn_3}
\ea
(We have again made use of the invariance under space--time translations
in $\langle \ldots \rangle_A$.)
We can parametrize the coordinates $b_i ~(i = 1, \ldots, n)$ as in Sect. 2,
i.e., in the form $b_i = b + p\tau_i ~(i = 1, \ldots, n)$, with a fixed $b$.
In fact, using $p_+ \simeq 2E$, $p_- \simeq 0$ and $\vec{p}_\perp \simeq
\vec{0}_\perp$, one has that $b_{i-} = 0$, $\vec{b}_{i\perp} = \vec{0}_\perp$
and $db_{i+} = 2E d\tau_i$. Eq. (\ref{asn_3}) becomes:
\ba
\lefteqn{
[ \langle \tilde{S}_{ij} (p,p'|A) \rangle_A ]_{(n)} \simeq } \nonumber \\
& & \simeq (2\pi)^4 \delta^{(4)} (q) ~{i \over p^2 - m^2 + i\varepsilon}
\displaystyle\int d\tau_1 \ldots \displaystyle\int d\tau_n
~\theta (\tau_n - \tau_{n-1}) \ldots \theta (\tau_2 - \tau_1)
\nonumber \\
& & \times \langle \{ \left[ -ig p^{\mu_n} A_{\mu_n} (b + p\tau_n) \right]
\ldots \left[ -ig p^{\mu_1} A_{\mu_1} (b + p\tau_1) \right] \}_{ij}
\rangle_A ~.
\label{asn_4}
\ea
In conclusion, summing all perturbative orders [Eqs. (\ref{asn_0}) and
(\ref{asn_4})], we find the following expression for
$\langle \tilde{S}_{ij} (p,p'|A) \rangle_A$:
\ba
\lefteqn{
\langle \tilde{S}_{ij} (p,p'|A) \rangle_A \simeq } \nonumber \\
& & \simeq (2\pi)^4 \delta^{(4)} (q) ~{i \over p^2 - m^2 + i\varepsilon}
\cdot \langle [W_p (b)]_{ij} \rangle_A \nonumber \\
& & = (2\pi)^4 \delta^{(4)} (q) ~{i \over p^2 - m^2 + i\varepsilon}
\cdot {\delta_{ij} \over N_c} \langle \Tr [W_p (b)] \rangle_A ~,
\ea
where $W_p (b)$ has been defined in Eq. (\ref{w_def}).
(In the last passage we have used the fact that the vacuum is
gauge--invariant.) By virtue of Eq. (\ref{as_prop}), we derive the following
expression for $\tilde{S}_{ij} (p)$ in the vicinity of the pole:
\be
\tilde{S}_{ij} (p) \mathop\simeq_{p^2 \to m^2}
{i \over p^2 - m^2 + i\varepsilon}
\cdot {\delta_{ij} \over N_c} \langle \Tr [W_p (b)] \rangle_A ~.
\label{as_pole_1}
\ee
Moreover, from the definition of the ``physical'' mass $m$ and of the residue
$Z_W$ at the pole for $p^2 \to m^2$ of the unrenormalized quark propagator,
we must have [see Eq. (\ref{s_mass})]:
\be
\tilde{S}_{ij} (p) \mathop\simeq_{p^2 \to m^2}
{i Z_W ~\delta_{ij} \over p^2 - m^2 + i\varepsilon} ~.
\label{as_pole_2}
\ee
From the comparison of Eqs. (\ref{as_pole_1}) and (\ref{as_pole_2}),
we derive the following expression for the residue $Z_W$ at the pole:
\be
Z_W = {1 \over N_c} \langle \Tr [W_p (b)] \rangle_A
= {1 \over N_c} \langle \Tr [W_p (0)] \rangle_A ~,
\ee
where we have again made use of the invariance under space--time translations.

\setcounter{subsection}{2}
\begin{flushleft}
{\large\bf \thesubsection The fermion propagator}
\end{flushleft}

\noindent
Let us consider now the unrenormalized full fermion propagator (not truncated!)
in the vicinity of the pole (i.e., for $p^2 \to m^2$). We start from the
fermion propagator $G_{ij} (x,y|A)$ in an external gluon field $A^\mu$
and then we take the functional average over the gluon field in order to get
the full propagator (\ref{f_prop}):
\be
\langle G_{ij} (x,y|A) \rangle_A = \langle T [ \psi_i(x) \overline{\psi}_j(y)
] \rangle = G_{ij} (x-y) ~.
\ee
Going to the momentum representation [see Eq. (\ref{fourier})], we obtain:
\be
\langle \tilde{G}_{ij} (p,p'|A) \rangle_A = (2\pi)^4 \delta^{(4)} (p' - p)
~\tilde{G}_{ij} (p) ~,
\ee
where $\tilde{G}_{ij} (p)$ is the unrenormalized propagator in the momentum
space, defined by Eq. (\ref{f_tilde}).
More precisely, we shall evaluate the following quantity:
\be
\overline{u}_\alpha (p') \langle \tilde{G}_{ij} (p,p'|A) \rangle_A u_\beta (p)
= (2\pi)^4 \delta^{(4)} (p' - p) ~\overline{u}_\beta (p) \tilde{G}_{ij} (p)
u_\alpha (p) ~,
\label{af_prop}
\ee
starting from the perturbative expansion of $\tilde{G}^{(tc)}_{ij} (p,p'|A)$
in the eikonal approximation, that we have computed in Sect. 3.  At the 0--th
order we find the following expression for the quantity (\ref{af_prop}):
\ba
\lefteqn{
[ \overline{u}_\alpha (p') \langle \tilde{G}_{ij} (p,p'|A) \rangle_A
u_\beta (p) ]_{(0)} = } \nonumber \\
& & = (2\pi)^4 \delta^{(4)} (p' - p)
~\overline{u}_\alpha (p') {i \over {\mathaccent 94 p} - m + i\varepsilon}
u_\beta (p) \nonumber \\
& & = (2\pi)^4 \delta^{(4)} (p' - p) ~4m^2 \delta_{\alpha\beta}
{i \over p^2 - m^2 + i\varepsilon} ~.
\label{afn_0}
\ea
At the $n$--th order ($n \ge 1$) we find, using the corresponding expression
for $[\tilde{G}^{(tc)}_{ij} (p,p'|A)]_{(n)}$ that we have derived in
Sect. 3 [see Eq. (\ref{fn_1})]:
\ba
\lefteqn{
[ \overline{u}_\alpha (p') \langle \tilde{G}_{ij} (p,p'|A) \rangle_A
u_\beta (p) ]_{(n)} = } \nonumber \\
& & = {i \over p'^2 - m^2 + i\varepsilon} \displaystyle\int {d^4 q_1 \over
(2\pi)^4} \ldots \displaystyle\int {d^4 q_n \over (2\pi)^4}
~(2\pi)^4 \delta^{(4)} (q - q_1 - \ldots - q_n) ~Q^{(ferm)}_{\alpha\beta,~ij}
(q_1, \ldots, q_n) \nonumber \\
& & \times {i \over (p + q_1 + \ldots + q_{n-1})^2 - m^2 + i\varepsilon}
\ldots {i \over (p + q_1)^2 - m^2 + i\varepsilon} \cdot
{i \over p^2 - m^2 + i\varepsilon} ~,
\label{afn_1}
\ea
where $q = p' - p$ and $Q^{(ferm)}_{\alpha\beta,~ij}
(q_1, \ldots, q_n)$ is given by:
\ba
\lefteqn{
Q^{(ferm)}_{\alpha\beta,~ij} (q_1, \ldots ,q_n) \equiv }
\nonumber \\
& & \equiv \langle \overline{u}_\alpha (p') ({\mathaccent 94 p}' + m)
\{ [ -i g \gamma^{\mu_n} \tilde{A}_{\mu_n} (q_n)
+ i \delta m (2\pi)^4 \delta^{(4)} (q_n) \cdot {\bf 1} ]
({\mathaccent 94 p} + {\mathaccent 94 q}_1 + \ldots +
{\mathaccent 94 q}_{n-1} + m) \nonumber \\
& & \ldots ({\mathaccent 94 p} + {\mathaccent 94 q}_1 + m)
[ -i g \gamma^{\mu_1} \tilde{A}_{\mu_1} (q_1)
+ i \delta m (2\pi)^4 \delta^{(4)} (q_1) \cdot {\bf 1} ] \}_{ij}
({\mathaccent 94 p} + m) u_\beta (p) \rangle_A \nonumber \\
& & = 4m^2 ~\langle N^{(ferm)}_{\alpha\beta,~ij} (q_1, \ldots ,q_n)
\rangle_A ~,
\ea
where $N^{(ferm)}_{\alpha\beta,~ij} (q_1, \ldots ,q_n)$ has been defined in
Eq. (\ref{f_num}). By using the result (\ref{f_num_fin}),
we find that the expression (\ref{afn_1}) simplifies as follows:
\ba
\lefteqn{
[ \overline{u}_\alpha (p') \langle \tilde{G}_{ij} (p,p'|A) \rangle_A
u_\beta (p) ]_{(n)} \simeq 4m^2 \delta_{\alpha\beta} \cdot
[ \langle [\tilde{S}_{ij} (p,p'|A) \rangle_A ]_{(n)} } \nonumber \\
& & \simeq 4m^2 \delta_{\alpha\beta} \cdot (2\pi)^4
\delta^{(4)} (q) ~{i \over p^2 - m^2 + i\varepsilon}
\displaystyle\int d\tau_1 \ldots \displaystyle\int d\tau_n
~\theta (\tau_n - \tau_{n-1}) \ldots \theta (\tau_2 - \tau_1)
\nonumber \\
& & \times \langle \{ \left[ -ig p^{\mu_n} A_{\mu_n} (b + p\tau_n) \right]
\ldots \left[ -ig p^{\mu_1} A_{\mu_1} (b + p\tau_1) \right] \}_{ij}
\rangle_A ~.
\label{afn_2}
\ea
In conclusion, summing all perturbative orders [Eqs. (\ref{afn_0}) and
(\ref{afn_2})], we obtain the following expression for the quantity
(\ref{af_prop}), evaluated in the vicinity of the pole:
\ba
\lefteqn{
\overline{u}_\alpha (p') \langle \tilde{G}_{ij} (p,p'|A) \rangle_A
u_\beta (p) \simeq 4m^2 \delta_{\alpha\beta} \cdot
\langle \tilde{S}_{ij} (p,p'|A) \rangle_A } \nonumber \\
& & \simeq 4m^2 \delta_{\alpha\beta} \cdot (2\pi)^4
\delta^{(4)} (q) ~{i \over p^2 - m^2 + i\varepsilon}
\cdot {\delta_{ij} \over N_c} \langle \Tr [W_p (b)] \rangle_A ~,
\ea
where $W_p (b)$ has been defined in Eq. (\ref{w_def}). (In the last passage
we have used the fact that the vacuum is gauge--invariant.) By virtue of Eq.
(\ref{af_prop}), we derive the following expression for
$\overline{u}_\alpha (p) \tilde{G}_{ij} (p) u_\beta (p)$
in the vicinity of the pole:
\be
\overline{u}_\alpha (p) \tilde{G}_{ij} (p) u_\beta (p)
\mathop\simeq_{p^2 \to m^2}
4m^2 \delta_{\alpha\beta} \cdot {i \over p^2 - m^2 + i\varepsilon}
\cdot {\delta_{ij} \over N_c} \langle \Tr [W_p (b)] \rangle_A ~.
\label{af_pole_1}
\ee
Moreover, from the definition of the ``physical'' mass $m$ and of the residue
$Z_W$ at the pole for $p^2 \to m^2$ of the unrenormalized quark propagator,
we must have [see Eq. (\ref{f_mass})]:
\be
\tilde{G}_{ij} (p) \mathop\simeq_{p^2 \to m^2}
{i Z_W ~\delta_{ij} \over {\mathaccent 94 p} - m + i\varepsilon}
= {i Z_W ~\delta_{ij} ~({\mathaccent 94 p} + m) \over p^2 - m^2
+ i\varepsilon} ~,
\ee
and, therefore:
\be
\overline{u}_\alpha (p) \tilde{G}_{ij} (p) u_\beta (p)
\mathop\simeq_{p^2 \to m^2}
4m^2 \delta_{\alpha\beta} \cdot {i Z_W ~\delta_{ij} \over p^2 - m^2
+ i\varepsilon} ~.
\label{af_pole_2}
\ee
From the comparison of Eqs. (\ref{af_pole_1}) and (\ref{af_pole_2}),
we derive the following expression for the residue $Z_W$ at the pole:
\be
Z_W = {1 \over N_c} \langle \Tr [W_p (b)] \rangle_A
= {1 \over N_c} \langle \Tr [W_p (0)] \rangle_A ~,
\label{af_res}
\ee
where we have made use of the invariance under space--time translations.
This expression is formally identical to the one we have obtained for the
scalar case: however, we must remember that now $\langle \ldots \rangle_A$
has to be intended as the functional average over the gluon field in
the theory with fermions. The same result (\ref{af_res}) was also derived
in Ref. \cite{Nachtmann91} using a different method.

\setcounter{subsection}{3}
\begin{flushleft}
{\large\bf \thesubsection The gluon propagator}
\end{flushleft}

\noindent
Let us consider, finally, the unrenormalized full gluon propagator
(not truncated!) in the vicinity of the pole (i.e., for $k^2 \to 0$).
We start from the ``gluon propagator'' $D^{a'a}_{\mu'\mu} (x,y|A)$ in an
external gluon field $A^\nu_b$ and then we take the functional average over
the gluon field $A^\nu_b$ in order to get the full propagator:
\be
\langle D^{a'a}_{\mu'\mu} (x,y|A) \rangle_A = \langle T [ A^{a'}_{\mu'} (x)
A^a_\mu (y) ] \rangle = D^{a'a}_{\mu'\mu} (x-y) ~.
\ee
Going to the momentum representation [see Eq. (\ref{fourier})], we obtain:
\be
\langle \tilde{D}^{a'a}_{\mu'\mu} (k,k'|A) \rangle_A
= (2\pi)^4 \delta^{(4)} (k' - k) ~\tilde{D}^{a'a}_{\mu'\mu} (k) ~,
\ee
where $\tilde{D}^{a'a}_{\mu'\mu} (k)$ is the full propagator in the momentum
space:
\be
\tilde{D}^{a'a}_{\mu'\mu} (k) \equiv \displaystyle\int d^4 z ~e^{ikz}
D^{a'a}_{\mu'\mu} (z) ~.
\ee
More precisely, we shall evaluate the following quantity:
\be
\varepsilon^{\mu'*}_{(\lambda')} (k') \langle \tilde{D}^{a'a}_{\mu'\mu}
(k,k'|A) \rangle_A \varepsilon^{\mu}_{(\lambda)} (k)
= (2\pi)^4 \delta^{(4)} (k' - k) ~\varepsilon^{\mu'*}_{(\lambda')} (k)
\tilde{D}^{a'a}_{\mu'\mu} (k) \varepsilon^{\mu}_{(\lambda)} (k) ~,
\label{ag_prop}
\ee
starting from the perturbative expansion of $\tilde{D}^{(tc)}_{\mu'\mu,~a'a}
(k,k'|A)$ in the eikonal approximation, that we have computed in Sect. 4.
At the 0--th order we find the following expression for the quantity
(\ref{ag_prop}):
\be
\left[ \varepsilon^{\mu'*}_{(\lambda')} (k') \langle
\tilde{D}^{a'a}_{\mu'\mu} (k,k'|A)
\rangle_A \varepsilon^{\mu}_{(\lambda)} (k) \right]_{(0)}
= (2\pi)^4 \delta^{(4)} (k' - k) ~\delta_{\lambda'\lambda} \delta_{a'a}
{i \over k^2 + i\varepsilon} ~.
\label{agn_0}
\ee
At the $n$--th order ($n \ge 1$) we find, using the corresponding expression
for the ``truncated--connected propagator'' $[\tilde{D}^{a'a}_{\mu'\mu}
(k,k'|A)]_{(n)}$ that we have derived in Sect. 4
[see Eqs. (\ref{gn_1}) -- (\ref{g_num_fin})]:
\ba
\lefteqn{
\left[ \varepsilon^{\mu'*}_{(\lambda')} (k') \langle
\tilde{D}^{a'a}_{\mu'\mu} (k,k'|A) \rangle_A
\varepsilon^{\mu}_{(\lambda)} (k) \right]_{(n)} \simeq }
\nonumber \\
& & \simeq \delta_{\lambda'\lambda}
{i \over k'^2 + i\varepsilon} \displaystyle\int {d^4 q_1 \over (2\pi)^4}
\ldots \displaystyle\int {d^4 q_n \over (2\pi)^4}
~(2\pi)^4 \delta^{(4)} (q - q_1 - \ldots - q_n) \nonumber \\
& & \times \langle \{ [-ig 2k^{\mu_n} \tilde{\cal A}_{\mu_n} (q_n)]
\ldots [-ig 2k^{\mu_1} \tilde{\cal A}_{\mu_1} (q_1)] \}_{a'a}
\rangle_A \nonumber \\
& & \times {i \over (k + q_1 + \ldots + q_{n-1})^2 + i\varepsilon}
\ldots {i \over (k + q_1)^2 + i\varepsilon} \cdot
{i \over k^2 + i\varepsilon} ~,
\label{agn_1}
\ea
where $q \equiv k' - k$ is the transferred momentum. This expression is
perfectly analogous to the corresponding expression obtained in Sect. A.1
for the scalar case. We can proceed as in that case to simplify Eq.
(\ref{agn_1}) as follows:
\ba
\lefteqn{
\left[ \varepsilon^{\mu'*}_{(\lambda')} (k') \langle
\tilde{D}^{a'a}_{\mu'\mu} (k,k'|A) \rangle_A
\varepsilon^{\mu}_{(\lambda)} (k) \right]_{(n)} \simeq }
\nonumber \\
& & \simeq (2\pi)^4 \delta^{(4)} (q)
~\delta_{\lambda'\lambda} ~{i \over k^2 + i\varepsilon}
\displaystyle\int d\tau_1 \ldots \displaystyle\int d\tau_n
~\theta (\tau_n - \tau_{n-1}) \ldots \theta (\tau_2 - \tau_1)
\nonumber \\
& & \times \langle \{ \left[ -ig k^{\mu_n} {\cal A}_{\mu_n} (b + k\tau_n)
\right] \ldots \left[ -ig k^{\mu_1} {\cal A}_{\mu_1} (b + k\tau_1) \right]
\}_{a'a} \rangle_A ~.
\label{agn_2}
\ea
In conclusion, summing all perturbative orders [Eqs. (\ref{agn_0}) and
(\ref{agn_2})], we find the following expression for the quantity
(\ref{ag_prop}), evaluated in the vicinity of the pole:
\be
\varepsilon^{\mu'*}_{(\lambda')} (k') \langle
\tilde{D}^{a'a}_{\mu'\mu} (k,k'|A) \rangle_A
\varepsilon^{\mu}_{(\lambda)} (k) \simeq
(2\pi)^4 \delta^{(4)} (q)
~\delta_{\lambda'\lambda} ~{i \over k^2 + i\varepsilon}
\cdot {\delta_{a'a} \over N_c^2 - 1} \langle \Tr [{\cal V}_k (b)] \rangle_A ~,
\ee
where ${\cal V}_k (b)$ has been defined in Eq. (\ref{v_def}) and we have
made use of the gauge--invariance of the vacuum (i.e., of the functional
average $\langle \ldots \rangle_A$). By virtue of Eq. (\ref{ag_prop}),
we derive the following expression for $\varepsilon^{\mu'*}_{(\lambda')} (k)
\tilde{D}^{a'a}_{\mu'\mu} (k) \varepsilon^{\mu}_{(\lambda)} (k)$
in the vicinity of the pole:
\be
\varepsilon^{\mu'*}_{(\lambda')} (k) \tilde{D}^{a'a}_{\mu'\mu} (k)
\varepsilon^{\mu}_{(\lambda)} (k) \mathop\simeq_{k^2 \to 0}
\delta_{\lambda'\lambda} ~{i \over k^2 + i\varepsilon}
\cdot {\delta_{a'a} \over N_c^2 - 1} \langle \Tr [{\cal V}_k (b)] \rangle_A ~.
\label{ag_pole_1}
\ee
Moreover, from the definition of the residue $Z_{\cal V}$ at the pole for
$k^2 \to 0$ of the unrenormalized full gluon propagator, we must have that:
\be
\tilde{D}^{a'a}_{\mu'\mu} (k) \mathop\simeq_{k^2 \to 0}
{-i Z_{\cal V} ~g_{\mu'\mu} \delta_{a'a} \over k^2 + i\varepsilon} ~,
\ee
and, therefore:
\be
\varepsilon^{\mu'*}_{(\lambda')} (k) \tilde{D}^{a'a}_{\mu'\mu} (k)
\varepsilon^{\mu}_{(\lambda)} (k) \mathop\simeq_{k^2 \to 0}
\delta_{\lambda'\lambda} \delta_{a'a} ~{i Z_{\cal V} \over k^2
+ i\varepsilon} ~.
\label{ag_pole_2}
\ee
From the comparison of Eqs. (\ref{ag_pole_1}) and (\ref{ag_pole_2}),
we derive the following expression for the residue $Z_{\cal V}$ at the pole:
\be
Z_{\cal V} = {1 \over N_c^2 - 1} \langle \Tr [{\cal V}_k (b)] \rangle_A
= {1 \over N_c^2 - 1} \langle \Tr [{\cal V}_k (0)] \rangle_A ~,
\ee
where we have made use of the invariance under space--time translations.

\vfill\eject

{\renewcommand{\Large}{\normalsize}
}

\vfill\eject

\noindent
\begin{center}
{\bf FIGURE CAPTIONS}
\end{center}
\vskip 0.5 cm
\begin{itemize}
\item [\bf Fig.~1.] a) The Feynman diagram corresponding to the $n$--th order
term ($n \ge 1$) in the perturbative expansion of the truncated--connected
quark propagator in an external gluon field $A^\mu_a$, in the eikonal
approximation.  [See Eq. (\ref{sn_1}) for the scalar case and Eqs.
(\ref{fn_1}) and (\ref{f_num_fin}) for the fermion case.]
b) The Feynman diagram which defines the $n$--th order perturbative term
($n \ge 1$) of the gluon matrix element (\ref{g_matrix}).
Crosses represent insertions of the external gluon field $A^\mu_a$. The
four--momenta $q_1, \ldots, q_n$ are taken to be flowing into the diagram.
\end{itemize}

\vfill\eject

\pagestyle{empty}

\centerline{\bf Figure 1}
\vskip 4truecm
\begin{figure}[htb]
\vskip 4.5truecm
\includegraphics{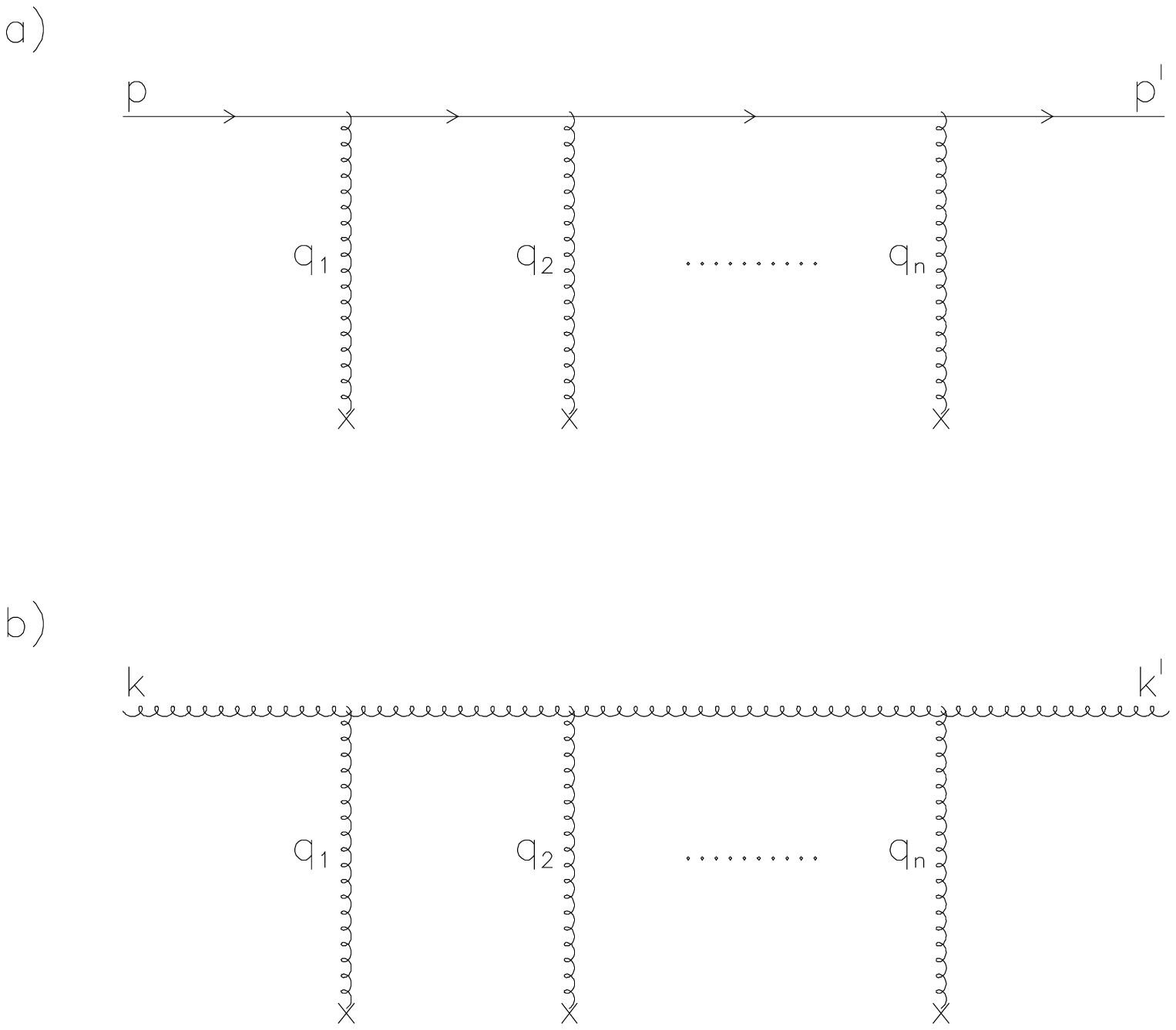}
\end{figure}

\vfill\eject

\end{document}